\documentclass{nature_dani}
\usepackage{amssymb,amsfonts,amsmath,rotating}
\usepackage{lscape}
\usepackage{hyperref}
\usepackage{graphicx}
\usepackage[font=small]{caption}
\usepackage{color}
\usepackage[linesnumbered,ruled,vlined]{algorithm2e}

\newcommand{\mc}{\mathcal}
\newcommand{\real}{\mathbb{R}}

\newcommand{\map}[3]{#1: #2 \rightarrow #3}
\newcommand{\transpose}{\mathsf{T}}

\title{Controllability Of Brain Networks}

\author{Shi Gu$^{1,2}$, Fabio Pasqualetti$^{3}$, Matthew Cieslak$^{4}$, Scott T. Grafton$^{4}$, Danielle S. Bassett$^{2,5,*}$}

\begin{document}

\maketitle

\begin{affiliations}
 \item Applied Mathematics and Computational Science Graduate Group, University of Pennsylvania, Philadelphia, PA, 19104, USA
 \item Complex Systems Group, Department of Bioengineering, University of Pennsylvania, Philadelphia, PA, 19104, USA
 \item Department of Mechanical Engineering, University of California, Riverside, CA, 92521, USA
 \item Department of Psychological and Brain Sciences and UCSB Brain Imaging Center, University of California, Santa Barbara, CA 93106, USA
 \item Department of Electrical Engineering, University of Pennsylvania, Philadelphia, PA, 19104, USA
\end{affiliations}

~\\
~\\
~\\
~\\
\begin{abstract}
Cognitive function is driven by dynamic interactions between large-scale neural circuits or networks, enabling behavior. Fundamental principles constraining these dynamic network processes have remained elusive. Here we use network control theory to offer a mechanistic explanation for how the brain moves between cognitive states drawn from the network organization of white matter microstructure. Our results suggest that densely connected areas, particularly in the default mode system, facilitate the movement of the brain to many easily-reachable states. Weakly connected areas, particularly in cognitive control systems, facilitate the movement of the brain to difficult-to-reach states. Areas located on the boundary between network communities, particularly in attentional control systems, facilitate the integration or segregation of diverse cognitive systems. Our results suggest that structural network differences between the cognitive circuits dictate their distinct roles in controlling dynamic trajectories of brain network function.
\end{abstract}




\newpage
\section*{Introduction}
Neuroscientific investigations seek to reveal how neural systems perform complex functions, how those systems and functions are altered in disease states, and how therapeutic interventions can be used to redirect these alterations. In essence, all three lines of investigation seek to address how neural systems move along dynamic trajectories: of cognitive function, disease, or recovery. Fundamental and therefore generalizable mechanisms of these movements have remained elusive.

The complexity of neural dynamics stems in part from the architectural complexity of the underlying anatomy. Different components (neurons, cortical columns, brain areas) are linked with one another in complex spatial patterns that enable diverse neural functions. These structural interactions can be represented as a graph or network, where component parts form the nodes, and where anatomical links form the edges between nodes. The architecture of these networks displays fascinating heterogenous features that play a role in neural function \cite{Bullmore2012}, development \cite{Feldt2011}, disease \cite{Bassett2009}, and sensitivity to rehabilitation \cite{Weiss2011}. Despite these recent discoveries, how these architectural features constrain neural dynamics in any of these phenomena is far from understood.

\begin{figure}[h]
\begin{center}
\includegraphics[width=3.5in]{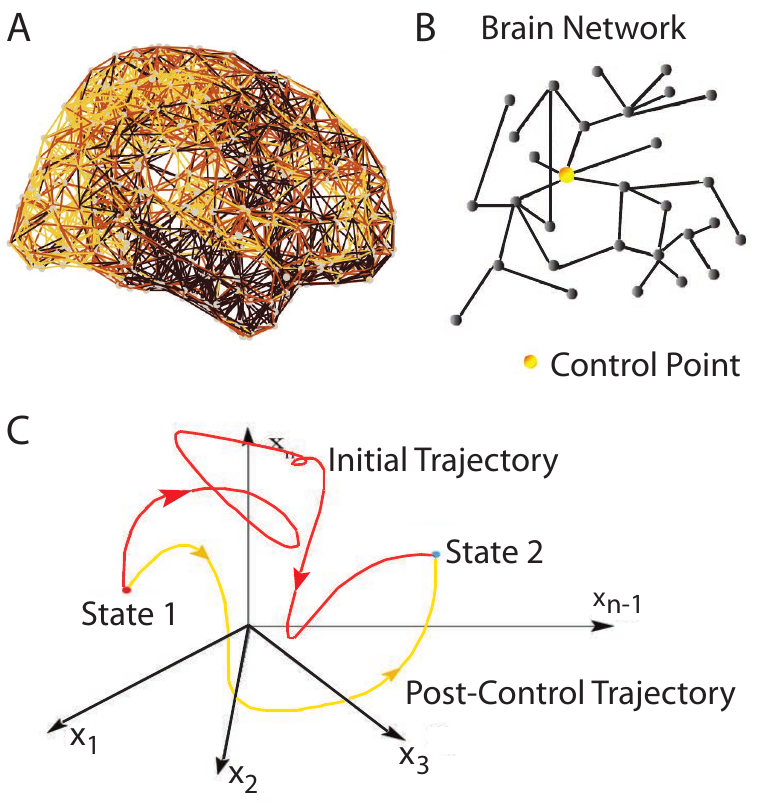}
\caption{\textbf{Conceptual Schematic.} From brain networks \emph{(A)}, we can estimate control points \emph{(B)} whose activity can move the brain into new dynamic trajectories (referred to as `post-control trajectories') that traverse diverse cognitive functions \emph{(C)}.}
 \label{fig0}
 \end{center}
 \end{figure}

Here we capitalize on recent theoretical advances in \emph{network control theory} to quantify the role of structural network organization on the dynamic trajectories of neural systems. Control of a network refers to the ability to manipulate local interactions of dynamic components to drive the global network along a desired trajectory, meaning a path traversing diverse system states. We postulate that the regulation of cognitive function is driven by a network-level control process akin to those utilized in other biological, technological, cyberphysical, and social systems. In this view, particular nodes (brain regions) at critical locations within the anatomical network topology act as drivers that are able to move the system (brain) into specific modes of action (cognitive functions). While not previously applied to neuroimaging data, network control theory provides a mathematical framework to investigate how structural features of a network impact controllability of cognitive dynamics.

We exploit network control theory to address two basic questions about how the large-scale circuitry of the human brain constrains its dynamics. First, is the human brain theoretically controllable? Based on the fact that cognitive, disease, and therapeutic processes can all alter the trajectories of brain function, we hypothesize that the brain is theoretically controllable. However, since each of these control processes requires alterations across distributed neural circuits, we conjecture that the brain is difficult to control via localized interventions. Second, which areas of the brain are most influential in constraining or facilitating changes in dynamic trajectories? We hypothesize that brain areas will play differential roles in these processes based on their topographical location within the wider network. Furthermore, because brain regions are organized into known cognitive systems, we postulate that structural differences between cognitive systems will predispose them to different roles in network control.

To address these questions, we predict the dynamic control properties of cognitive function based on independent \emph{structural} properties of the brain. We build structural brain networks from diffusion spectrum imaging (DSI) data acquired in triplicate from $8$ healthy human adults. We perform diffusion tractography to estimate the number of streamlines linking $N=234$ large-scale cortical and subcortical regions extracted from the Lausanne atlas \cite{Hagmann2008}. We summarize these estimates in a weighted adjacency matrix whose entries reflect the number of streamlines connecting different regions. This construction enables us to examine brain network controllability at both the global and regional levels in individual participants.

\section*{Mathematical Models}

\subsection{Dynamic Model of Neural Processes}
Neural activity evolves through neural circuits as a collection of dynamic processes. These processes can be approximated by linearized generalizations \cite{Galan2008} of nonlinear models of cortical circuit activity \cite{Honey2009}. To study the control properties of neural processes, we focus on a simplified noise-free linear discrete-time and time-invariant network model:
\begin{equation}\label{eq: linear network}
  \mathbf{x} (t+1) = \mathbf{A} \mathbf{x}(t) + \mathbf{B}_{\mc K} \mathbf{u}_{\mc K} (t) ,
\end{equation}
where $\map{\mathbf{x}}{\real_{\ge 0}}{\real^N}$ describes the state (i.e., electrical charge or oxygen level) of brain regions over time, and $\mathbf{A} \in \real^{N \times N}$ is
a symmetric and weighted adjacency matrix whose elements $A_{ij}$
indicate the number of streamlines connecting region $i$ and region
$j$, which we scale to ensure stability of the dynamic process at long time intervals \cite{Horn1985}. Note that $A_{ii}=0$. The input matrix $\mathbf{B}_{\mc K}$ identifies the control points in the brain $\mc K = \{k_1, \dots, k_m \}$, where \begin{align}\label{eq: B}
  B_{\mc K} =
  \begin{bmatrix}
    e_{k_1} & \cdots & e_{k_m}
  \end{bmatrix},
\end{align}
and $e_i$ denotes the $i$-th canonical vector of dimension $N$. The input $\map{\mathbf{u}_{\mc K}}{\real_{\ge  0}}{\real^m}$ denotes the control strategy.

\subsection{Network Controllability} The notion of controllability of
a dynamical system introduced by Kalman et al. \cite{REK-YCH-SKN:63} refers to the possibility of driving the state of a dynamical system to a specific target state by means of a control input. Classic results in control theory ensure that controllability of the network
\eqref{eq: linear network} from the set of network nodes $\mc K$ is
equivalent to the controllability Gramian $\mathbf{W}_{\mc K}$ being
invertible, where \begin{equation}
  \mathbf{W}_{\mathcal{K}} = \sum_{\tau =0}^{\infty}\mathbf{A}^\tau
  \mathbf{B}_{\mathcal{K}}\mathbf{B}_{\mathcal{K}}^\transpose \mathbf{A}^\tau .
\end{equation}
The eigenvalues of the controllability Gramian are a quantitative
measure of the degree of controllability of different network
configurations and trajectories. The structure of the Gramian itself
can be further used to provide systematic guidelines for the selection
of control areas in the brain that can optimize different cognitive
functions, as we demonstrate in the following section.

\section*{Results}

\subsection{Global Controllability}
We first sought to address the question: ``Is the human brain
theoretically controllable?''. To answer this question, we computed
the eigenvalues of the controllability Gramian for each brain region as control node, and for each of the 24 diffusion
imaging scans. We observed that the smallest eigenvalues were
consistently greater than $0$, which demonstrates that the system is
theoretically controllable through a single brain region. While
estimated to be nonzero, however, the values of the smallest
eigenvalues were extremely small (mean $2.5\times 10^{-23}$, STD
$4.8\times 10^{-23}$) with respect to the largest
  eigenvalues (always greater or equal to $1$), indicating that in
practice the system is extremely hard to control with interventions
localized to any single brain region.

\subsection{Regional Controllability}
We next sought to address the question: ``Which areas of the brain are most influential in constraining or facilitating changes in dynamic trajectories?''. To address this question, we employ $3$ diagnostics of regional controllability developed in network control theory: the average, modal, and boundary controllability. Each of these diagnostics captures a different control goal \cite{FP-SZ-FB:13q}. Average controllability identifies brain areas that can steer the system into many easily reachable states, that is, states that are reachable with limited input energy. Modal controllability identifies brain areas that can steer the system into difficult-to-reach states, that is, states requiring a substantial control effort. Boundary controllability identifies brain areas that can steer the system into states where different cognitive systems are either decoupled or integrated. For mathematical definitions of these diagnostics, see Materials and Methods.

\begin{figure}
 \centerline{\includegraphics[width=4in]{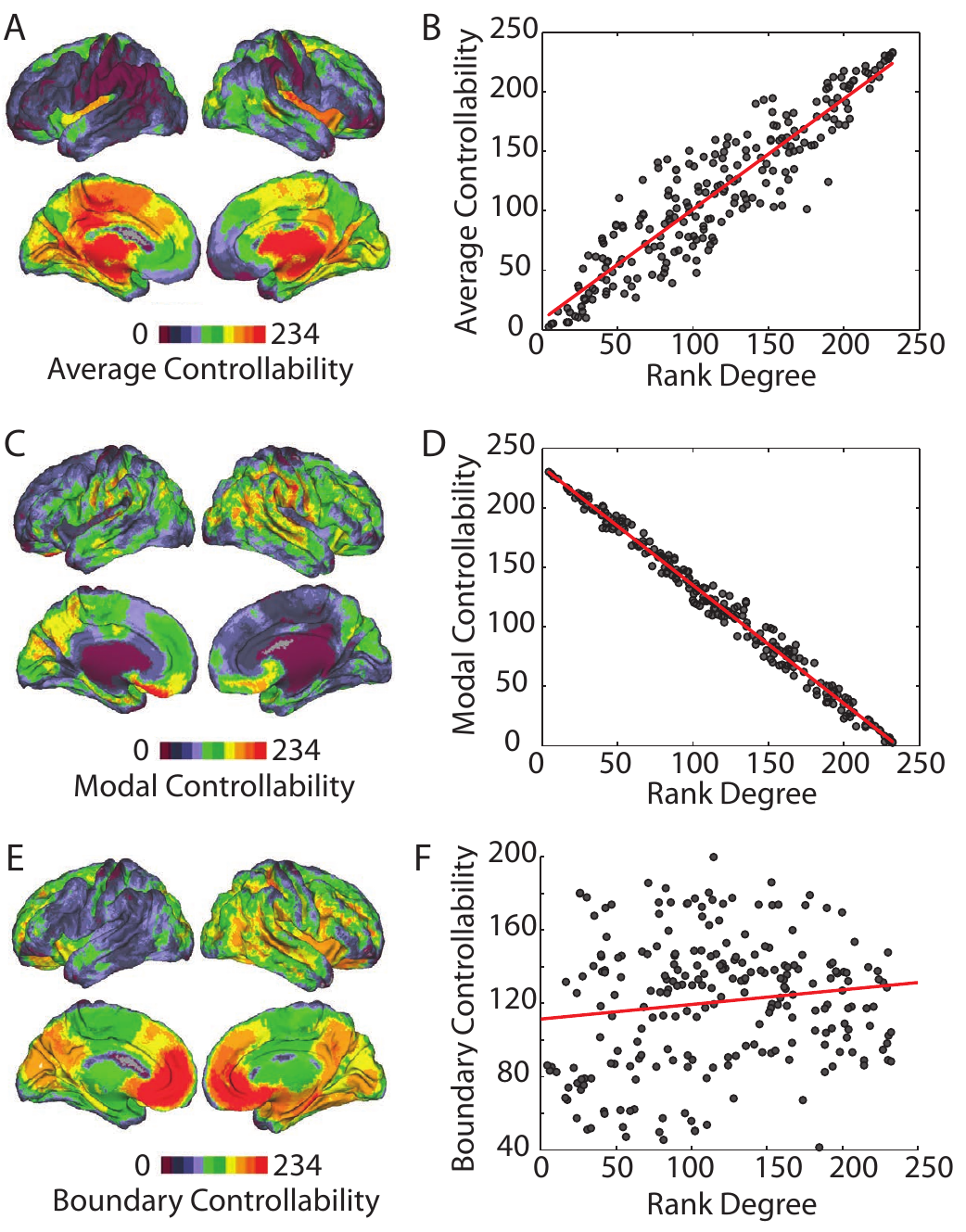}}
\caption{\textbf{Brain Network Control Properties} \emph{(A)} Average controllability quantifies control to many states. Here we show average controllability values ranked for all brain regions plotted on a surface visualization. Warmer colors indicate larger values of average controllability. \emph{(B)} Scatter plot of degree (ranked for all brain regions) versus average controllability (Pearson correlation $r=0.91$, $p=8\times 10^{-92}$). \emph{(C)} Modal controllability quantifies control to difficult-to-reach states. Here we show modal controllability values ranked for all brain regions plotted on a surface visualization. \emph{(D)} Scatter plot of degree (ranked for all brain regions) versus modal controllability ($r=-0.99$, $p=2\times 10^{-213}$). \emph{(E)} Boundary controllability quantifies control to decouple or integrate network modules. Here we show boundary controllability values ranked for all brain regions plotted on a surface visualization. \emph{(F)} Scatter plot of degree (ranked for all brain regions) versus boundary controllability ($r = 0.13$, $p = 0.03$). In panels \emph{(A)}, \emph{(C)}, and \emph{(E)}, warmer colors indicate larger controllability values, which have been averaged over both replicates and subjects. These results are reliable over a range of atlas resolutions (see the SI). \label{con}}
 \end{figure}

\paragraph{Average Controllability}
Average controllability identifies brain areas that can steer the system into many different states. The average controllability is greatest in precuneus, posterior cingulate, superior frontal, paracentral, precentral and subcortical structures (Fig.~\ref{con}A). These areas are strikingly similar to those reported to be the structural ``core'' of the human cerebral cortex \cite{Hagmann2008}, which innervate the rest of the network with a high density of connections. In other words, these regions are ``hubs'', having high network degree, which is defined as the number of edges emanating from that region. To validate this relationship, we show that the average controllability of all brain regions is strongly correlated with degree (Pearson correlation $r=0.91$, $p=8\times 10^{-92}$; Fig.~\ref{con}B).

\paragraph{Modal Controllability}
Modal controllability identifies brain areas that can steer the system into difficult-to-reach states. The modal controllability is greatest in postcentral, supramarginal, inferior parietal, pars orbitalis, medial orbitofrontal, and rostral middle frontal cortices (Fig.~\ref{con}C). In contrast to areas with high average controllability, areas with high modal controllability are not hubs of the network but instead have low degree. The modal controllability of all brain regions is strongly anti-correlated with degree (Pearson correlation $r=-0.99$, $p=2\times 10^{-213}$; Fig.~\ref{con}D). The inverse relationship between degree and modal controllability is consistent with the notion that difficult-to-reach states require the control of poorly connected areas.

\paragraph{Boundary Controllability}
Boundary controllability identifies brain areas that can steer the system into states where different cognitive systems are either decoupled or isolated. This control goal complements but differs from those of average and modal controllability. The boundary controllability is greatest in rostral middle frontal, lateral orbitofrontal, frontal pole, medial orbitofrontal, superior frontal, and anterior cingulate cortices (Fig.~\ref{con}E). In contrast to areas with high average or modal controllability, areas with high boundary controllability are neither hubs nor non-hubs. The boundary controllability of all brain regions is not strongly correlated or strongly anti-correlated with degree (Pearson correlation $r=0.13$, $p=0.03$; Fig.~\ref{con}F).

\begin{figure}
 \centerline{\includegraphics[width=6in]{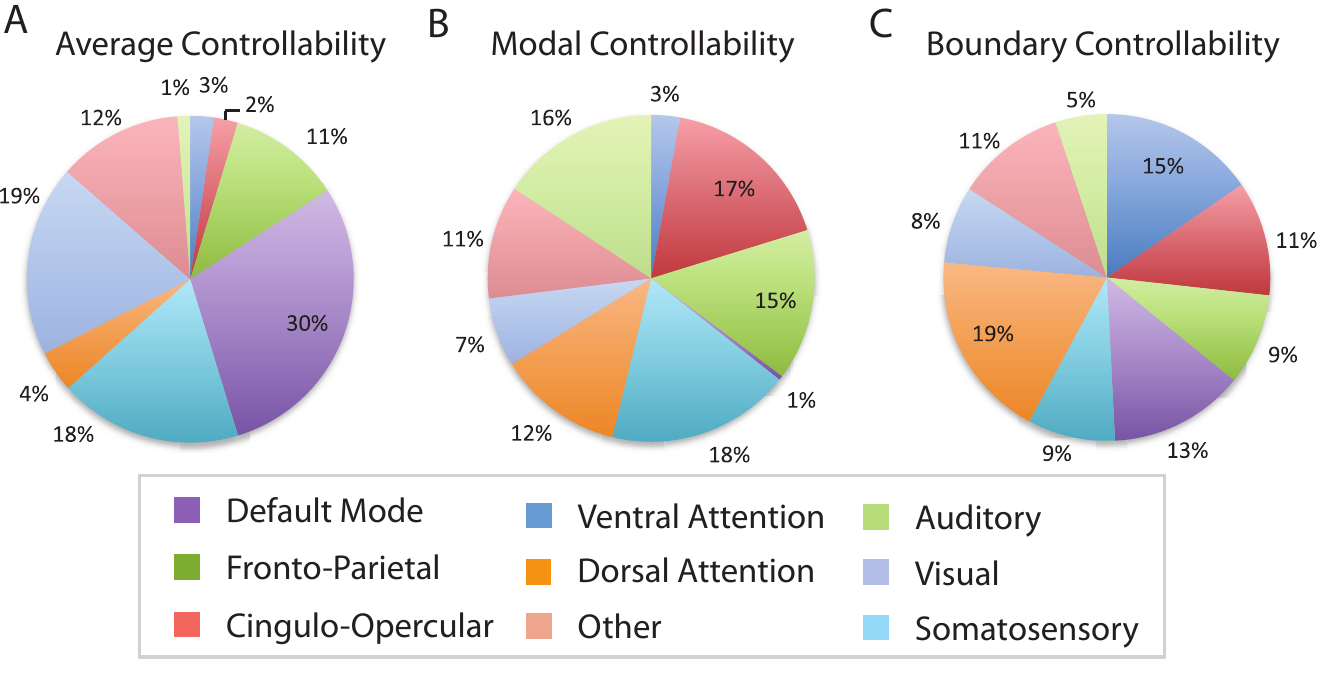}}
\caption{\textbf{Control Roles of Cognitive Systems} Cognitive control hubs are differentially located across cognitive systems. \emph{(A)} Hubs of average controllability are preferentially located in the default mode system. \emph{(B)} Hubs of modal controllability are predominantly located in cognitive control systems, including both the fronto-parietal and cingulo-opercular systems. \emph{(C)} Hubs of boundary controllability are distributed throughout all systems, with the two predominant systems being ventral and dorsal attention systems. Control hubs have been identified at the group level as the 30 regions with the highest controllability values (averaged over replicates and subjects). \label{pie}}
 \end{figure}

\subsection{Regional Controllability of Cognitive Systems}
Finally, we asked the question ``Are control regions differentially located in or between known cognitive systems?''. Drawing from the literature, we formulate 3 specific hypotheses addressing this question. First, based on the fact that average controllability identifies areas of the brain that may be important in steering the system into many easily reachable states, we hypothesize that areas of high average controllability would map on to areas active in the brain's baseline or ``default'' state (the resting state), from which the brain smoothly moves to multitudinous task states. In contrast, modal controllability identifies areas of the brain that may be important in steering the system to difficult-to-reach states. We hypothesize that areas of high modal controllability would therefore map on to areas responsible for the brain's transitions between difficult tasks, specifically executive areas involved in cognitive control. Finally, boundary controllability identifies areas of the brain that can steer the system into states where different cognitive systems are either decoupled or integrated. Because these areas mathematically sit at the boundaries between network communities or putative functional modules, we expect that these areas would map relatively uniformly onto all cognitive systems: each system having a few boundary nodes that might play a role in linking that system to another. However, we also postulate a particular enrichment of the attention systems, based on their role in feature selection, gating, orienting and multi-tasking which constrain integration across other cognitive systems.

To test these hypotheses, we assigned the 234 regions of the Lausanne atlas to the following large-scale cortical networks, which we refer to as ``cognitive systems'': auditory, visual, sensorimotor, ventral attention, dorsal attention, default mode, fronto-parietal, and cingulo-opercular. This set of cognitive systems, and the association of regions to these cognitive systems, has previously been extracted from resting state data using a network-based clustering approach \cite{Power2012} and has been widely applied to examine the roles of cognitive systems in task-based and resting-state connectivity \cite{Cole2013,Power2013,Cole2014} (see the SI for regional attributions to systems).

We find that regions of high controllability are differentially associated with the 8 cognitive systems (Fig.~\ref{pie}). We define the set of high control hubs as the 30 regions with the largest controllability values (averaged over all scans), and we calculate the percent of hubs present from each of the 8 cognitive systems. To correct for system size, we normalize the percentage by the number of regions in a cognitive system. Consistent with our hypotheses, 30\% of average control hubs lie in the default mode system, 32\% of modal control hubs lie in the fronto-parietal and cingulo-opercular cognitive control systems, and 36\% of boundary control hubs lie in the ventral and dorsal attention systems. Our results are qualitatively similar if we choose a larger or smaller set of control hubs (see the SI).
 \begin{figure} [h]
 \centerline{\includegraphics[width=4in]{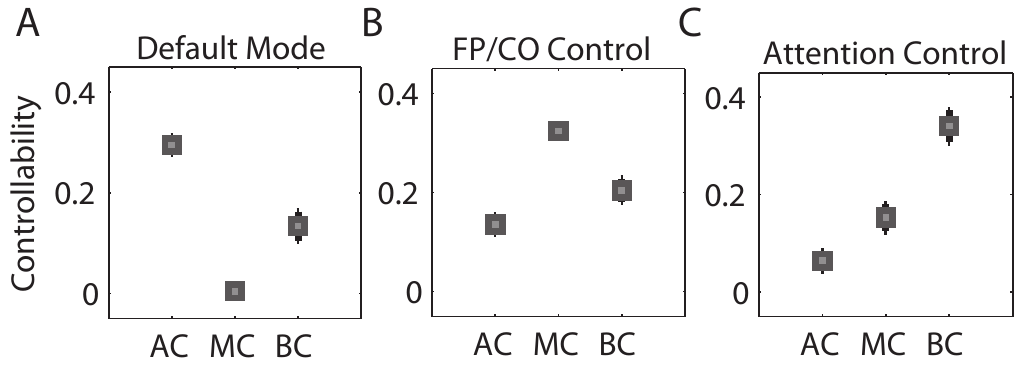}}
\caption{\textbf{Differential Recruitment of Cognitive Systems to Network Control} Average controllability (AC), modal controllability (MC), and boundary controllability (BC) hubs are differentially located in default mode \emph{(A)}, fronto-parietal and cingulo-opercular cognitive control \emph{(B)}, and attentional control \emph{(C)} systems. Values are averaged over the 3 replicates for each individual; error bars indicate standard deviation of the mean over subjects.  \label{dsys}}
 \end{figure}
These results suggest the presence of a controllability-by-system interaction: certain types of controllability may be utilized or enabled by different cognitive systems. To directly test for this interaction, we extract control hubs for each scan, determine their association with the three hypothesized control systems (default mode, fronto-parietal and cingulo-opercular cognitive control, and attentional control), and quantify the mean controllability value for all hubs in each system (Fig~\ref{dsys}). We observe that regions of the default mode system form strong average controllability hubs but weaker modal and boundary controllability hubs. Regions of the cognitive control networks (fronto-parietal and cingulo-opercular) form strong modal controllability hubs and regions of the attentional control networks (ventral and dorsal) form strong boundary controllability hubs. To statistically validate this finding, we perform a repeated measures 2-way Analysis of Variance with cognitive system and controllability diagnostic as categorical factors, and with scan replicate as a repeated measure. The main effect of system is significant ($F(9)=42.40$; $p=0$), the main effect of diagnostic is significant ($F(2)=22.25$, $p=0.0013$), and the interaction between system and diagnostic is also significant ($F(18)=39.81$; $p=0$). These statistics indeed suggest that structural differences between the default mode, cognitive control, and attentional control systems may facilitate their distinct roles in controlling dynamic trajectories of brain network function.

\section*{Discussion}

The brain is a networked dynamical system that moves between diverse cognitive states to enable complex behaviors. Fundamental principles constraining these dynamic trajectories have remained elusive. Here we use network control theory to offer a mechanistic explanation for how the brain moves between cognitive states drawn from the network organization of white matter microstructure. Our results indicate that densely connected areas are theoretically expected to facilitate the movement of the brain to many easily-reachable states and we show that these areas are preferentially located in the default mode system. Weakly connected areas, predominantly located in cognitive control systems, are theoretically expected to facilitate the movement of the brain to difficult-to-reach states. Finally, areas located on the boundary between network communities, predominantly located in attentional control systems, are theoretically expected to facilitate the integration or segregation of diverse cognitive systems. As a whole, this body of work suggests that structural network differences between the default mode, cognitive control, and attentional control systems dictate their distinct roles in controlling dynamic trajectories of brain network function.

\subsection{Theoretically Predicted Controllability of Large-Scale Neural Circuitry}

The relationship between any mathematical measure of controllability, and what it means for a brain to be in control is unknown. Nevertheless, network controllability diagnostics provide theoretical predictions regarding the controllability of large-scale neural circuitry. Using the smallest eigenvalues of the controllability Gramian, we show that structural brain network architecture is theoretically controllable, but pragmatically very difficult to control. The theoretical possibility of controlling the brain from a single region is consistent with a large body of scientific evidence stemming from (i) patient studies that demonstrate that lesions to single brain areas can have dramatic effects on regional activity, inter-regional connectivity, and by extension cognitive function and behavior \cite{Grefkes2014}, and (ii) real-time fMRI studies of neuromodulation that demonstrate that subjects can control the activity of single brain regions to modulate pain perception \cite{deCharms2005}. These findings are consistent with the theoretical expectation that it is possible -- with a single input -- to move the brain to a single target state. However, our prediction that the brain is pragmatically difficult to control indicates that it is practically impossible to move the brain to \emph{any} target state that we might desire with little control action. This predicted difficulty is consistent with the fact that even complex combinations of drugs, brain stimulation, and cognitive therapies \cite{Pallanti2014} can still fail to right cognitive function when it has gone wrong. These findings suggest that some dynamic trajectories (from disease to health for example) are extremely difficult and practically impossible, particularly from single inputs to single brain areas. The complexity of cognitive function and its underlying mechanisms, illustrated by control difficulty, calls for new tools to quantify and understand which trajectories are amenable to control, thereby informing the use of targeted therapies including brain stimulation \cite{Johnson2013}.

\subsection{The Role of Hubs in Brain Control}

We can study brain controllability either globally, as described above, or locally. Network control theory posits the diagnostic of average controllability as a quantification of a node's role in moving the system to many easily reachable states. We show that brain regions with high average controllability tend to be areas with a large number of white matter streamlines connecting them to the rest of the network -- that is, network hubs. These hubs tend to be located in areas of the default mode system. This suggests that the brain has a baseline, resting state organization which is optimized to allow the brain to move to a large number of easily reachable states. Regions of high average controllability, which have the greatest predicted influence in moving the brain to this plethora of states, are highly active at rest in the default mode network. If we assume that the brain has been optimized over evolutionary time scales to maximally enable a complex functional battery \cite{Bassett2010c,Bullmore2012}, these results suggest the tantalizing possibility that the large majority of complex functions performed by the brain are easily reachable from the default mode state. The few functions which might be difficult to reach from the default mode state may utilize alternative control mechanisms, including modal and boundary control.

The fact that structural hubs, particularly in the default mode network, play such a striking role in brain network controllability may help to explain the growing body of evidence indicating that disease states can preferentially target hub areas \cite{Bassett2009}. \emph{In silico} studies suggest that lesions to highly structurally connected areas have a greater impact on ensuing functional connectivity than lesions to sparsely connected areas \cite{Alstott2009}. Moreover, alterations to default mode hubs are associated with drastic changes in cognitive function associated with normative aging \cite{Tomasi2012} and neurodegenerative disorders like Alzheimer's \cite{Buckner2009}. Our results provide a mechanistic explanation for these findings by suggesting that hubs form key control points in brain networks; alterations to hub regions can therefore have disproportionately high impacts on system function.

\subsection{The Role of Weak Connections in Brain Control}

While our results demonstrate that hubs are theoretically implicated in moving the brain to many easily reachable states, weakly connected areas are critical for moving the brain to difficult-to-reach states. We observe that these modal control points, while distributed across the brain, tend to be predominantly located in cognitive control systems including the fronto-parietal and cingulo-opercular networks. These two systems are characterized by different functional connectivity patterns at rest \cite{Power2012} and are thought to support distinct functional roles within the general area of cognitive control \cite{Elton2014}: task-switching \cite{Stoet2009} and task-set maintenance \cite{Shenhav2013}, respectively. Our results suggest a fundamental underlying mechanism of cognitive control: brain regions sparsely interconnected with the rest of the brain are critically important for moving the system into difficult-to-reach states. This theoretical hypothesis is consistent with the increased engagement of the cognitive control system in highly effortful tasks \cite{Lifshitz2013}.

More generally, the fact that weak connections play a critical role in system dynamics is one that has traditionally received little attention \cite{Schneidman2006,Granovetter1983}. However, recent work has begun to demonstrate the relevance of weak connections for both cognitive function and psychiatric disease. For example, the topology of weak connections in resting state fMRI could be used to classify healthy volunteers versus schizophrenia patients, while the topology of strong connections could not \cite{Bassett2012}. Moreover, in healthy individuals, the topology of weak connections more accurately correlates with intelligent quotients than the topology of strong connections \cite{Cole2012,Santarnecchi2014}. These findings challenge the traditional view of a prominent role of strong connections in brain dynamics. Our results provide a mechanistic rationale for the importance of weak connections, which are theoretically critical in enabling a system to move to difficult-to-reach states, which may include high performance states (such as measured by IQ) or altered performance states (such as those present in psychiatric conditions).

\subsection{The Role of Community Structure in Brain Control}

In addition to the two mechanisms that enable trajectories to (i) many easily reachable states, and (ii) a few difficult to reach states, networked systems often utilize a third mechanism -- boundary controllability -- which enables the segregation or integration of network modules. Modular structure has been reported in structural \cite{Chen2008}, functional \cite{Meunier2009}, and dynamic \cite{Bassett2011b} brain networks. In resting state connectivity studies, these modules have been linked to known cognitive systems \cite{Power2012}. Our results suggest that a widely distributed set of brain areas across all of these systems enables segregation and integration of putative cognitive modules. We also observe a particular enrichment of boundary control hubs in dorsal and ventral attentional systems, suggesting that attentional control may be implemented by boundary control strategies integrating or segregating disparate cognitive systems. Such a theoretical prediction is supported by evidence that attentional control integrates different cognitive functions \cite{Corbetta2002,Pessoa2009}, and that disconnections of attentional networks is accompanied by extensive cognitive deficits \cite{Corbetta2011,Parks2013}.

\subsection{Methodological Considerations}
The controllability diagnostics that we report and utilize here are highly reliable across multiple scanning sessions (see SI), indicating their potential use in explaining individual differences in cortical function. Moreover, the anatomical distribution of controllability diagnostics is consistent across 5 parcellation schemes segregating the brain into 83, 129, 234, 463, and 1015 regions of interest (see SI), suggesting that these measures are robust quantifications of brain dynamics.

\subsection{Conclusion}
A fundamental understanding of the principles by which the brain transitions between diverse cognitive states enabling behavior would necessarily have far-reaching implications for basic cognitive neuroscience and applications in myriad clinical domains. Our results suggest that structural design could underlie basic cognitive control processes, via the fundamental mechanism of network controllability. These findings lay the groundwork for future studies examining relationships between individual differences in network controllability diagnostics and behavioral, cognitive, clinical, and genetic variables.

\section*{Methods}

\subsection{Data Acquisition and Preprocessing}
Diffusion spectrum images (DSI \cite{Wedeen2005}) were acquired for a total of $8$ subjects in triplicate (mean age $27\pm5$ years, $2$ female, $2$ left handed) along with a $T1$ weighted anatomical scan at each scanning session \cite{Cieslak2014}. DSI scans sampled $257$ directions using a $Q5$ half shell acquisition scheme with a maximum $b$ value of $5000$ and an isotropic voxel size of $2.4$mm. We utilized an axial acquisition with the following parameters: $TR=11.4$s, $TE=138$ms, $51$ slices, FoV ($231$,$231$,$123$ mm). All participants volunteered with informed consent in accordance with the Institutional Review Board/Human Subjects Committee, University of California, Santa Barbara.

DSI data were reconstructed in DSI Studio (www.dsi-studio.labsolver.org) using $q$-space diffeomorphic reconstruction (QSDR) \cite{Yeh2011}. QSDR first reconstructs diffusion weighted images in native space and computes the quantitative anisotropy (QA) in each voxel. These QA values are used to warp the brain to a template QA volume in MNI space using the SPM nonlinear registration algorithm. Once in MNI space, spin density functions were again reconstructed with a mean diffusion distance of $1.25$mm using three fiber orientations per voxel. Fiber tracking was performed in DSI Studio with an angular cutoff of $55^{\circ}$, step size of $1.0$mm, minimum length of $10$mm, spin density function  smoothing of $0.0$, maximum length of $400$mm and a QA threshold determined by DWI signal in the CSF. Deterministic fiber tracking using a modified FACT algorithm was performed until $100,000$ streamlines were reconstructed for each individual.

Anatomical scans were segmented using FreeSurfer \cite{Dale1999} and parcellated according to the Lausanne 2008 atlas included in the connectome mapping toolkit \cite{Gerhard2011,Hagmann2008}. A parcellation scheme including 234 regions was registered to the B0 volume from each subject's DSI data. The B0 to MNI voxel mapping produced via QSDR was used to map region labels from native space to MNI coordinates. To extend region labels through the gray/white matter interface, the atlas was dilated by $4$mm. Dilation was accomplished by filling non-labeled voxels with the statistical mode of their neighbors' labels. In the event of a tie, one of the modes was arbitrarily selected. Each streamline was labeled according to its terminal region pair.

\subsection{Network Controllability Diagnostics} We examine 3 diagnostics of controllability utilized in the network control literature: \emph{average controllability}, \emph{modal controllability}, and \emph{boundary controllability}.

\noindent \textbf{Average Controllability} Average
  controllability of a network -- formally defined as $\text{Trace}
  (\mathbf{W}_{\mc K}^{-1})$ -- equals the average input energy from a
  set of control nodes and over all possible target states
  \cite{BM-DK-DG:04,HRS-MT:12}. Motivated by the relation
  $\text{Trace} ( \mathbf{W}_{\mc K}^{-1}) \ge N^2 / \text{Trace} (
  \mathbf{W}_{\mc K})$, recent results in the control of networked
  systems \cite{THS-JL:13}, and the fact that $\mathbf{W}_{\mc K}$ is
  close to singularity, we adopt $\text{Trace} ( \mathbf{W}_{\mc K})$
  as a measure of the average controllability of a network. Regions
  with high average controllability are most influential in the
  control of network dynamics over all different target states.

\noindent \textbf{Modal Controllability} Modal
  controllability refers to the ability of a node to control each
  evolutionary mode of a dynamical network \cite{AMAH-AHN:89}, and can
  be used to identify the \emph{least controllable} state from a set
  of control nodes. Modal controllability is computed from the
  eigenvector matrix $V = [v_{ij}]$ of the network adjacency matrix
  $\mathbf{A}$. By extension from the PBH test \cite{TK:80}, if the
  entry $v_{ij}$ is small, then the $j$-th mode is poorly controllable
  from node $i$. Following \cite{FP-SZ-FB:13q}, we define $\phi_i =
  \sum_{j =1}^N (1 - \lambda_j^2 (A)) v_{ij}^2$ as a scaled measure of
  the controllability of all $N$ modes $\lambda_1 (A),\dots, \lambda_N
  (A)$ from the brain region $i$. Regions with high modal
  controllability are able to control all the dynamic modes of the
  network, and hence to drive the dynamics towards
  hard-to-reach configurations.

\noindent \textbf{Boundary Controllability} Boundary
  controllability measures the ability of a set of control nodes to
  decouple the dynamic trajectories of disjoint brain regions. To
  evaluate the boundary controllability of different brain regions, we
  proceed as follows. First, we compute a \emph{robust partition} of
  the brain network as described in \cite{DSB-MAP-NFW-STG-JMCPJM:13},
  and we identify the set of $N_1$ boundary nodes. We assign to these boundary nodes the
  boundary controllability value of $1$. Second, following
  \cite{FP-SZ-FB:13q}, we determine the two-partition of the least controllable subnetwork from its Fiedler eigenvector \cite{SF:10,MF:73}, and we identify the additional boundary nodes.
  We assign to these boundary nodes the boundary controllability value
  of $(N-N_1)/N$. Finally, we iterate this process until all nodes
  have been assigned a boundary controllability value.

Average, modal, and boundary controllability each provide a scalar
value for each brain region. To enable direct comparison between
controllability diagnostics and across different subjects, we perform
ranking and normalization steps. In particular, for each of the
controllability diagnostics we (i) rank the scalar values for each
subject, and (ii) average the ranked values across the subjects.

\section*{Acknowledgments}
DSB acknowledges support from the Alfred P. Sloan Foundation, the Army Research Laboratory through contract no. W911NF-10-2-0022 from the U.S. Army Research Office, the Institute for Translational Medicine and Therapeutics at Penn, and the National Science Foundation award \#BCS-1441502. SG acknowledges support from the Applied Mathematics and Computational Science Graduate Program at the University of Pennsylvania. MC and STG were supported by PHS Grant NS44393 and the Institute for Collaborative Biotechnologies through grant W911NF-09-0001 from the U.S. Army Research Office. The content is solely the responsibility of the authors and does not necessarily represent the official views of any of the funding agencies.

\clearpage
\newpage
\newpage

\section*{Supplemental Results}

In the main manuscript, we utilize a parcellation of the cortical and subcortical tissue into $N=234$ different brain regions. This parcellation is in fact part of a wider family of Lausanne atlas parcellations that include the following:
\begin{itemize}
\item Scale 33: $N=83$ brain regions
\item Scale 60: $N=129$ brain regions
\item Scale 125: $N=234$ brain regions
\item Scale 250: $N=463$ brain regions
\item Scale 500: $N=1015$ brain regions
\end{itemize}
This multi-scale atlas has previously been used to examine the hierarchical nature of brain network topography \cite{Hagmann2008}. In this supplement, we examine the reproducibility of our results obtained using Scale 125 (described in the main manuscript) across the remaining spatial resolutions provided by the other 4 atlases.

\subsection{Global Controllability Across Spatial Scales}
\addcontentsline{toc}{subsection}{Global Controllability Across Spatial Scales}

We calculated the global controllability of each node in each atlas for each person and scan. We observed that the mean global controllability (averaged across subjects, scans, and nodes) decreases with the spatial scale of the atlas: see Table\ref{TabGC}. Note: here we report the mean and STD of global controllability diagnostics over brain regions.

\newpage
\begin{table*}
\begin{center}
\caption[Global Controllability Diagnostics]{Global Controllability Diagnostic Values (GC) over the 5 Scales of the Lausanne Atlas Family.}
\label{TabGC}
\begin{tabular} {rccc}
Scale & Number of Nodes & mean GC & STD GC\\
\hline
33 & 83 & $2.55 \times 10^{-21}$ & $1.61 \times 10^{-21}$ \\
60 & 129 & $5.78 \times 10^{-22}$ & $3.88 \times 10^{-22}$ \\
125 & 234 & $4.52 \times 10^{-23}$ & $3.59 \times 10^{-23}$ \\
250 & 463 & $7.10 \times 10^{-25}$ & $7.65 \times 10^{-25}$ \\
500 & 1015 & $2.09 \times 10^{-27}$ & $7.23 \times 10^{-27}$ \\
\hline
\end{tabular}
\end{center}
\end{table*}

\subsection{Reproducibility of Controllability Diagnostics Across Spatial Scales}
\addcontentsline{toc}{subsection}{Reproducibility of Controllability Diagnostics Across Spatial Scales}

\begin{figure}
 \centerline{\includegraphics[width=6in]{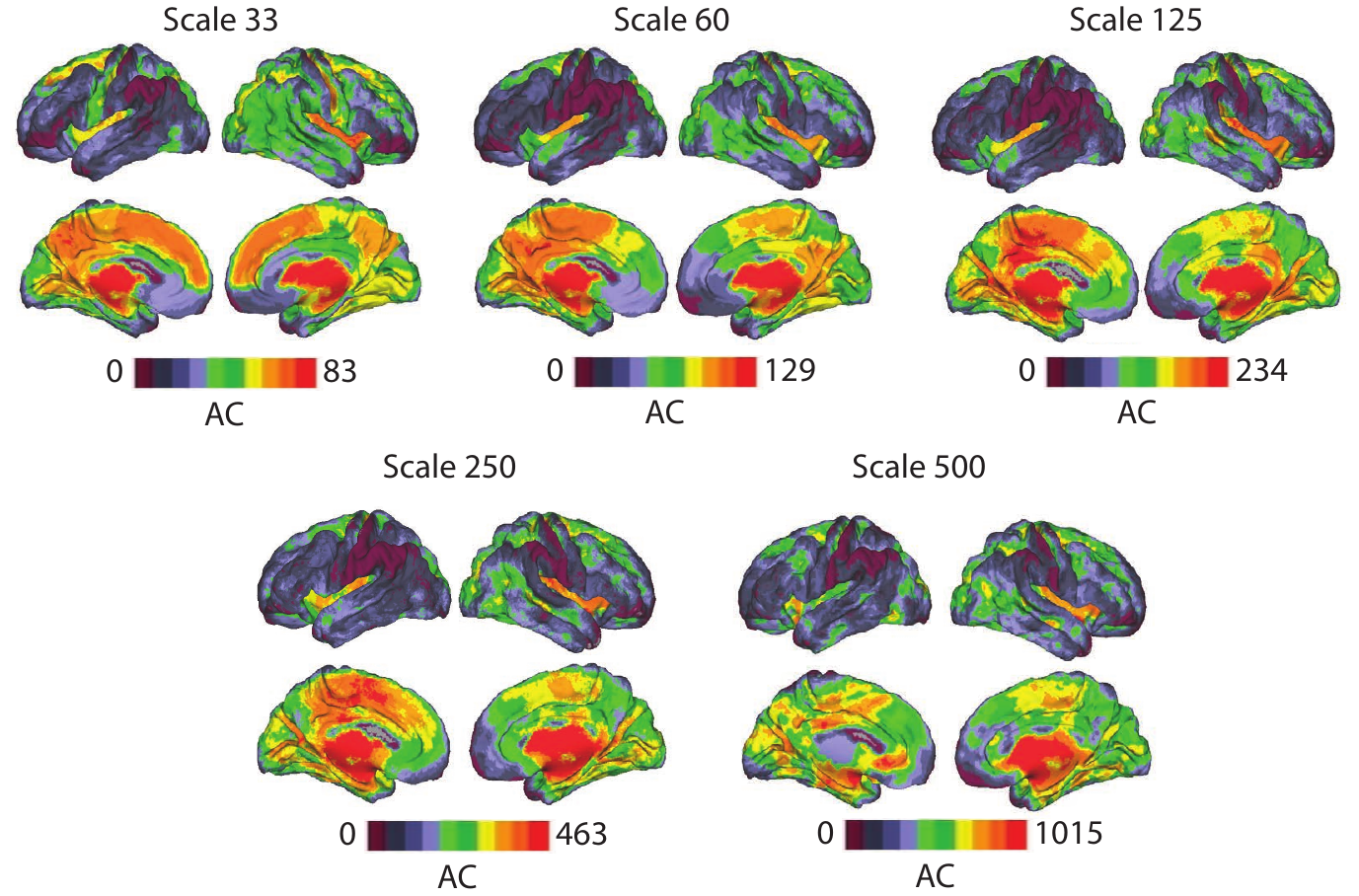}}
\caption[Average Controllability Across Spatial Scales]{\textbf{Average Controllability Across Spatial Scales} Surface visualizations of the ranked average controllability (AC) values over the 5 spatial scales of the Lausanne atlas \cite{Hagmann2008}. \label{aver}}
 \end{figure}

\begin{figure}
 \centerline{\includegraphics[width=6in]{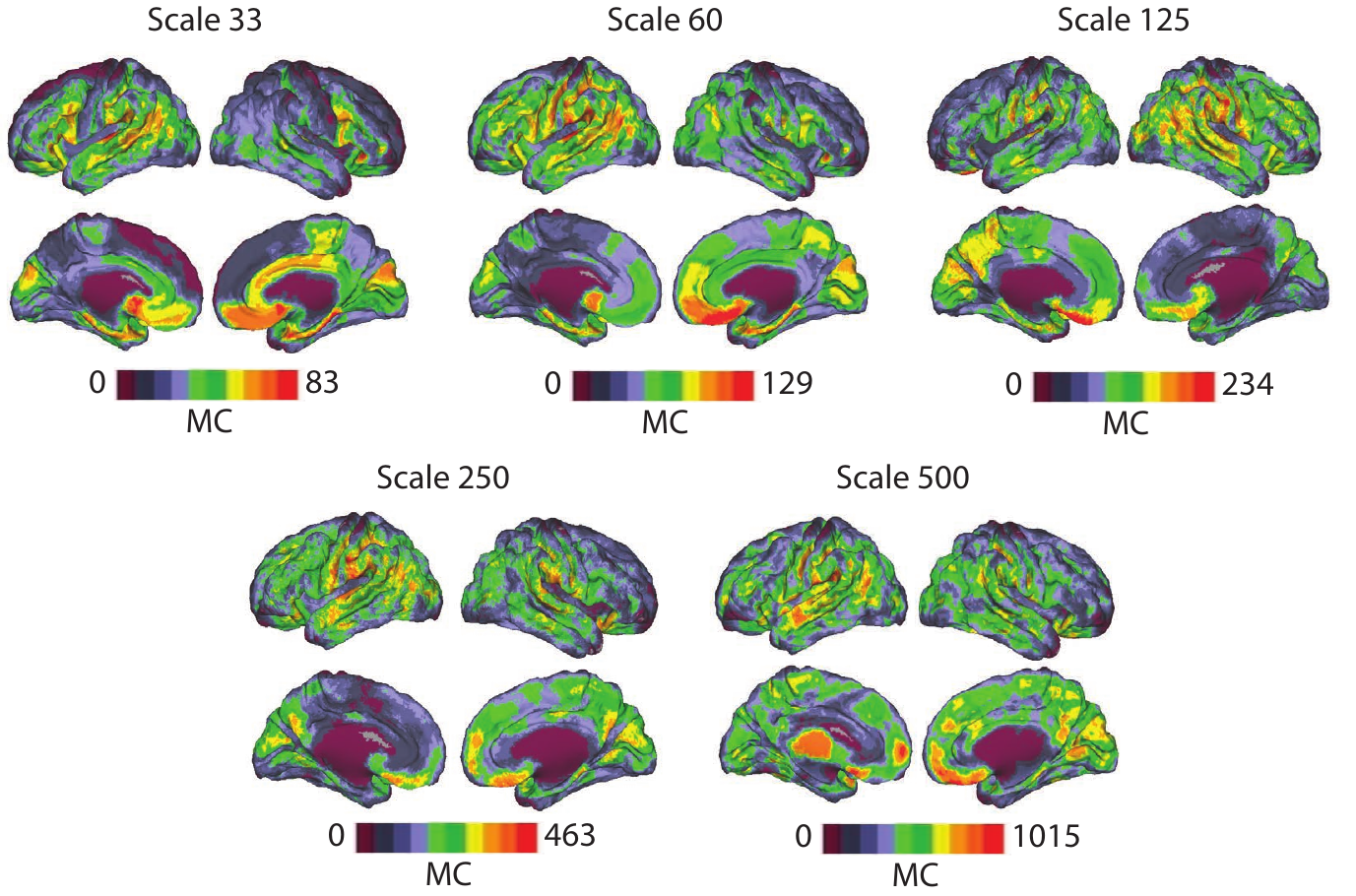}}
\caption[Modal Controllability Across Spatial Scales]{\textbf{Modal Controllability Across Spatial Scales} Surface visualizations of the ranked modal controllability (MC) values over the 5 spatial scales of the Lausanne atlas \cite{Hagmann2008}. \label{mod}}
 \end{figure}

\begin{figure}
 \centerline{\includegraphics[width=6in]{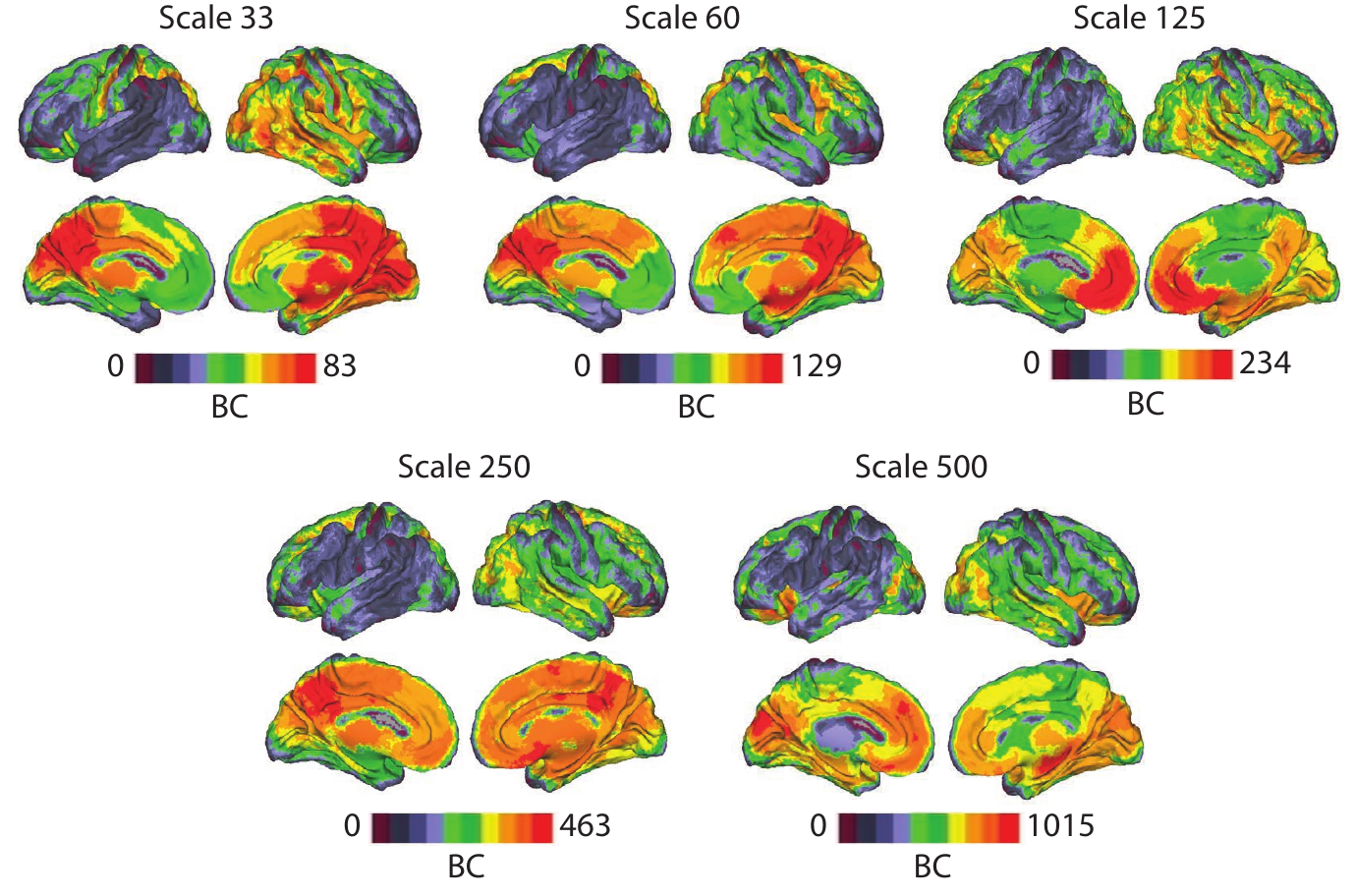}}
\caption[Boundary Controllability Across Spatial Scales]{\textbf{Boundary Controllability Across Spatial Scales} Surface visualizations of the ranked boundary controllability (BC) values over the 5 spatial scales of the Lausanne atlas \cite{Hagmann2008}. \label{boun}}
 \end{figure}

In the main manuscript, we show the anatomical distribution of the 3 controllability diagnostics over the $N=234$ brain regions of the Scale 125 atlas. In Fig.~\ref{aver} of this supplement, we show that the anatomical distribution of average controllability is visually similar across all 5 spatial scales assessed with the entire Lausanne atlas family. In Fig.~\ref{mod} and Fig.~\ref{boun}, we show a similar reproducibility of the anatomical distribution of modal and boundary controllability, respectively.

\subsection{Reproducibility of Degree-Controllability Correlations Across Spatial Scales}
\addcontentsline{toc}{subsection}{Reproducibility of Degree-Controllability Correlations Across Spatial Scales}

In the main manuscript, we observed that for Scale 125 ($N=234$) the degree was strongly positively correlated with the average controllability, strongly negatively correlated with the modal controllability, and neither strongly positively nor strongly negatively correlated with the boundary controllability. In Table~\ref{Tab1}, we report the correlations between degree and the 3 controllability diagnostics as a function of spatial resolution: from Scale 33 ($N=83$) to Scale 500 ($N=1015$). We observe that the degree-controllability correlations reported for Scale 125 are reproducibly observed across the remaining 4 spatial scales, comprising both higher and lower spatial resolutions.

\begin{table*}
\begin{center}
\caption[Correlation between Degree and Controllability Diagnostics]{Pearson correlation coefficients $r$ between rank degree, average controllability (AC), boundary controllability (BC), and modal controllability (MC).}
\label{Tab1}
\begin{tabular} {rccccc}
~ & ~ & Degree & AC & BC & MC\\
\hline
\textbf{Scale 33} & ~ & ~ & ~ & ~ & ~\\
\hline
~ &Degree                        &        1.0000        &     0.9764             &   0.5225             &  -0.9923 \\
~ &AC &      0.9764          &        1.0000              &    0.6302          &       -0.9688 \\
~ &BC &      0.5225     &               0.6302        &	1.0000	&  -0.5120                   \\
~ &MC  &     -0.9923           &    -0.9688                   &     -0.5120       & 1.0000 \\
\hline
\textbf{Scale 60} & ~ & ~ & ~ & ~ & ~\\
\hline
~ &Degree     &        1.0000        &     0.9429             &   0.4733                  &        -0.9912         \\
~ &AC &      0.9429          &        1.0000        &    0.6262          &       -0.9320   \\
~ &BC &      0.4733     &               0.6252        &	1.0000	&  -0.4806                   \\
~ &MC  &     -0.9912           &    -0.9320                   &     -0.4806       & 1.0000 \\
\hline
\textbf{Scale 125} & ~ & ~ & ~ & ~ & ~\\
\hline
~ &Degree        &        1.0000        &     0.9205             &   0.1385                  &        -0.9937         \\
~ &AC &      0.9205          &        1.0000   &    0.1461          &       -0.9125           \\
~ &BC &      0.1385     &               0.1461        &	1.0000	&  -0.1270                   \\
~ &MC  &     -0.9937           &    -0.9125                   &     -0.1270       & 1.0000 \\
\hline
\textbf{Scale 250} & ~ & ~ & ~ & ~ & ~\\
\hline
~ &Degree   &        1.0000        &     0.9114             &   0.3310                  &        -0.8626         \\
~ &AC &      0.9114          &        1.0000  &    0.4785          &       -0.7822           \\
~ &BC &      0.3310     &               0.4785        &	1.0000	&  -0.2968                   \\
~ &MC  &     -0.8626           &    -0.7822            &     -0.2968       & 1.0000 \\
\hline
\textbf{Scale 500} & ~ & ~ & ~ & ~ & ~\\
\hline
~ &Degree   &        1.0000        &     0.9122             &   0.2709                  &        -0.9962         \\
~ &AC &      0.9122          &        1.0000       &    0.3637   &       -0.9042           \\
~ &BC &      0.2709     &               0.3637        &	1.0000	&  -0.2366       \\
~ &MC  &     -0.9962           &    -0.9042                   &     -0.2366       & 1.0000 \\
\end{tabular}
\end{center}
\end{table*}

\subsection{Test-Retest Reliability of Controllability Diagnostics}
\addcontentsline{toc}{subsection}{Test-Retest Reliability of Controllability Diagnostics}

When proposing a new diagnostic of brain network architecture, it is critical to determine the reliability of those diagnostic values across iterative measurement. Here we capitalize on the fact that the same 8 subjects whose data are reported in the main manuscript were imaged over 3 different days. We utilize these iterative scans to assess the test-retest reliability of the 3 controllability diagnostics.

To compare the results among different scans and subjects, we consider the average correlation. Suppose we have $n$ subjects and for each of them we have $K$ scans with corresponding controllability values $c^i_1, \cdots, c^i_K$. The averaged correlation between controllability diagnostic values for subject $i$ and subject $j$ is defined as
\begin{equation}
R^{B}_{ij} = \frac{\sum_{s=1}^{K}\sum_{t=1}^{K} \mathrm{corr}(c^i_s, c^j_t) }{K^2}
\end{equation}
for subject $i\neq j$ and where $s$ and $t$ index scanning sessions, and $\mathrm{corr}$ indicates the calculation of a Pearson correlation coefficient. The average correlation between controllability diagnostic values for the same subject across scanning sessions is defined as
\begin{equation}
R^{W}_{ii} =  \frac{\sum_{s\neq t} \mathrm{corr}(c^i_s, c^i_t)}{K(K-1)}
\end{equation}
for $i = j$.
We refer to the quantity $R^{B}_{ij}$ as the average between-subject correlation and to the quantity $R^{W}_{ii}$ as the average within-subject correlation.

We report the within- and between-subject correlations for all 3 controllability diagnostics and for global controllability across all 5 spatial scales of the Lausanne atlas family in Tab.~\ref{Tab2}. We observe that all 3 controllability diagnostics display significantly greater within-subject correlation than between-subject correlation, indicating that these diagnostics are statistically reproducible across scanning sessions and significantly different across individuals. The average and modal controllability display a relatively high mean $R$ (approximately $0.90$) and relatively low standard error. While still statistically reproducible across scanning sessions, the boundary controllability displays a lower mean $R$ than the average and modal controllability, and a higher standard error. The global controllability is not reproducible across scanning sessions. These observations are consistently observed across the 5 spatial scales of the Lausanne atlas family of parcellations.

\begin{table*}
\begin{center}
 \caption[Test-Retest Reliability of Controllability Diagnostics]{Test-Retest Reliability of Controllability Diagnostics: average controllability (AC), boundary controllability (BC), modal controllability (MC) and global controllability (GC).}
 \label{Tab2}
 \begin{tabular}{cccccc}
   ~ & ~ & AC & BC & MC & GC\\
    \hline
     \textbf{Scale 33} & ~ & ~ & ~ & ~ & ~ \\
    \hline
   ~ &   Mean Within  &   0.9642        & 0.7250 &   0.9708  & 0.0674\\
   ~ &   Mean Between &   0.8966         & 0.3436 &  0.9191  & 0.0501\\
   ~ &   STE Within &  	  0.0222         & 0.1279 &  0.0119  & 0.0527\\
   ~ &   STE Between &   0.0227        & 0.2014 &  0.0168  & 0.0426\\
   ~ &   $p$-value             & 5.7e-11          & 2.5e-6   & 9.2e-12 & 0.4626 \\
   \hline
    \textbf{Scale 60} & ~ & ~ & ~ & ~ & ~\\
    \hline
    ~ & Mean Within  &   0.9510        & 0.6146 &   0.9519 & 0.0662 \\
    ~ &Mean Between &   0.8449         & 0.3465   &  0.8544  & 0.0501\\
    ~ &STE Within &  	  0.0283         & 0.1552 &  0.0199  & 0.0590\\
    ~ &STE Between &   0.0333        & 0.1351 & 0.0261   & 0.0387\\
    ~ &$p$-value             &   4.0e-12        & 2.8e-6   & 9.5e-15 & 0.3076\\
    \hline
    \textbf{Scale 125} & ~ & ~ & ~ & ~&~ \\
    \hline
    ~ & Mean Within  &   0.9404        & 0.5147 &   0.9348 & 0.0782\\
    ~ &Mean Between &   0.8036         & 0.1954 &  0.7900  & 0.0527\\
    ~ &STE Within &  	  0.0298         & 0.1311 &  0.0234  & 0.0508\\
    ~ &STE Between &   0.0383        & 0.1638 & 0.0350  & 0.0449 \\
    ~ &$p$-value             & 5.3$\times 10^{-14}$    & 1.8$\times 10^{-6}$   & 1.1$\times 10^{-16}$ & 0.1442\\
     \hline
    \textbf{Scale 250} & ~ & ~ & ~ & ~ &~\\
    \hline
    ~ & Mean Within  &   0.9320        & 0.5192 &   0.9230 & 0.0481\\
    ~ &Mean Between &   0.7751         & 0.2208 &  0.7451  & 0.0391\\
    ~ &STE Within &  	  0.0256         & 0.2012 &  0.0227  & 0.0383\\
    ~ &STE Between &   0.0332        & 0.2077 & 0.0267  & 0.0286\\
    ~ & $p$-value             &    4.5e-19       & 4.1e-6   & 7.8e-26& 0.4268 \\
      \hline
    \textbf{Scale 500} & ~ & ~ & ~ & ~ &~\\
    \hline
    ~ & Mean Within  &   0.9090        & 0.4990 &   0.8982 & 0.0395\\
    ~ & Mean Between &   0.7261         & 0.2367 &  0.6909  & 0.0229\\
    ~ & STE Within  &  	  0.0280         & 0.1373 &  0.0263  & 0.0327\\
    ~ & STE Between &   0.0338        & 0.1290 & 0.0224  & 0.0241\\
    ~ &$p$-value             &  1.0e-21          & 1.4e-6  & 4.6e-33  & 0.0864 \\
 \end{tabular}
 \end{center}
 \end{table*}

\subsection{Reproducibility of Control Roles of Cognitive Systems}

In the main text, we observed that a 30\% of average control hubs lie in the default mode system, 32\% of modal control hubs lie in the fronto-parietal and cingulo-opercular cognitive control systems, and 34\% of boundary control hubs lie in the ventral and dorsal attention systems. Here we demonstrate that these results are qualitatively reproduced for different definitions of control hubs: namely, the 25 nodes with the highest control values (out of a possible 234 nodes), the 30 nodes with the highest control values (as shown in the main manuscript), or the 35 nodes with the highest control values. When control hubs are defined as the top 25 nodes, we observe that 32\% of average control hubs lie in the default mode system, 33\% of modal control hubs lie in the fronto-parietal and cingulo-opercular systems, and 33\% of boundary control hubs lie in the ventral and dorsal attentional systems. When control hubs are defined as the top 35 nodes, we observe that 28\% of average control hubs lie in the default mode system, 31\% of modal control hubs lie in the fronto-parietal and cingulo-opercular systems, and 32\% of boundary control hubs lie in the ventral and dorsal attentional systems. These results demonstrate that the presence of a controllability-by-system interaction is robust to small variation in the choice of the size of the control hub set.

\begin{figure*}
\begin{center}
\includegraphics[width=5in]{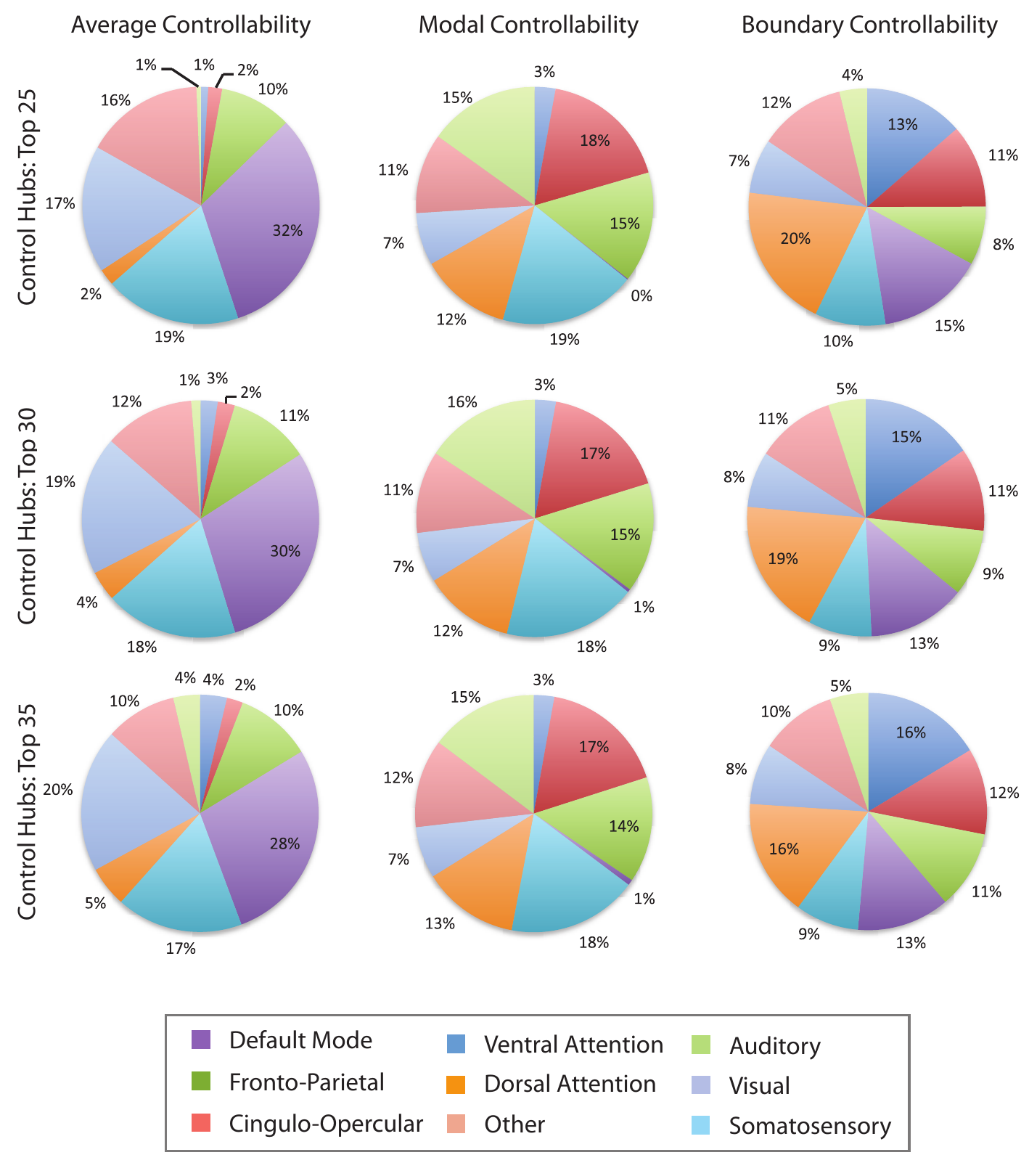}
\caption[Control Roles of Cognitive Systems]{\textbf{Control Roles of Cognitive Systems.} Cognitive control hubs are differentially located across cognitive systems. \emph{(Left)} Hubs of average controllability are preferentially located in the default mode system. \emph{(Middle)} Hubs of modal controllability are predominantly located in cognitive control systems, including both the fronto-parietal and cingulo-opercular systems. \emph{(Right)} Hubs of boundary controllability are distributed throughout all systems, with the two predominant systems being ventral and dorsal attention systems. These anatomical distributions are consistent across different definitions of control hubs as either \emph{(Top)} the 25 nodes with the highest control values, \emph{(Middle)} the top 30 nodes with the highest control values, or \emph{(Bottom)} the 35 nodes with the highest control values.}
 \label{pie_supmat}
 \end{center}
 \end{figure*}

\subsection{Reproducibility of Differential Recruitment of Cognitive Systems to Network Control}

\begin{figure*}
\begin{center}
\includegraphics[width=5in]{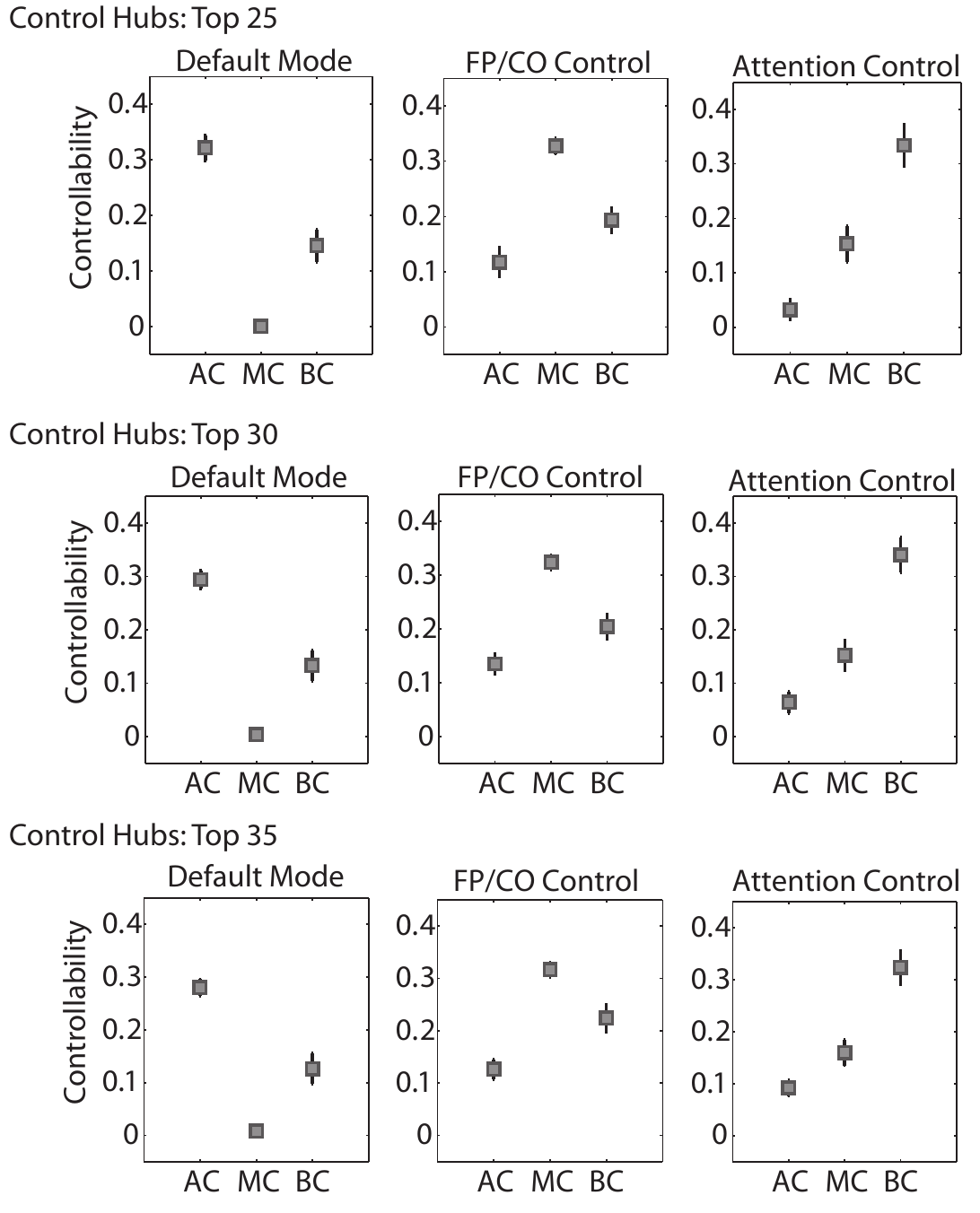}
\caption[Differential Recruitment of Cognitive Systems to Network Control]{\textbf{Differential Recruitment of Cognitive Systems to Network Control.} Average controllability (AC), modal controllability (MC), and boundary controllability (BC) hubs are differentially located in default mode \emph{(A)}, fronto-parietal and cingulo-opercular cognitive control \emph{(B)}, and attentional control \emph{(C)} systems. These results are consistently observed whether we define control hubs as the 25 nodes with the highest control values \emph{(Top Row)}, the 30 nodes with the highest control values \emph{(Middle Row)}, or the 35 nodes with the highest control values \emph{(Bottom Row)}. Values are averaged over the 3 replicates for each individual; error bars indicate standard deviation of the mean over subjects.}
 \label{diff_recruit_supmat}
 \end{center}
 \end{figure*}

In the main text, we observed the presence of a controllability-by-system interaction, and interpreted this as indicative of the possibility that certain types of controllability may be utilized or enabled by different cognitive systems. In particular, we observed that regions of the default mode system form strong average controllability hubs but weaker modal and boundary controllability hubs. Regions of the cognitive control networks (fronto-parietal and cingulo-opercular) form strong modal controllability hubs and regions of the attentional control networks (ventral and dorsal) form strong boundary controllability hubs. Here we demonstrate that these results are qualitatively reproduced for different definitions of control hubs: namely, the 25 nodes with the highest control values, the 30 nodes with the highest control values (as shown in the main manuscript), or the 35 nodes with the highest control values; see Fig.~\ref{diff_recruit_supmat}.

In the main text, we validate this finding by performing a repeated measures 2-way Analysis of Variance (ANOVA) with system and controllability diagnostic as categorical factors, and with scan replicate as a repeated measure. Here, we performed the same ANOVA for the case in which control nodes are defined as the 25 nodes with the highest control values or the 35 nodes with the highest control values, and found similar results in both cases: (i) for top 25 nodes, the main effect of system is $F(9)=43.7716$ ($p=0$), the main effect of diagnostic is $F(2)=16.5413$ ($p=2.0553e-4$), and the interaction between system and diagnostic is $F(18)=42.1475$ ($p=0$); (ii) for the top 35 nodes, the main effect of system is $F(9)=34.3787$ ($p=0$), the main effect of diagnostic is $F(2)=9.7420$ ($p=0.0022$), and the interaction between system and diagnostic is $F(18)=36.9762$ ($p=0$). Consistent with the results reported in the main manuscript, these statistics indeed suggest that structural differences between the default mode, cognitive control, and attentional control systems may facilitate their distinct roles in controlling dynamic trajectories of brain network function. We observe that the weaker control hubs we include in the analysis (i.e., larger number of control hubs), the less significant the relationship to cognitive systems. This suggests that the strong control hubs are significantly associated with cognitive systems but that weak control nodes may not be.

\subsection{Robustness of Results to Alternative Weighting Schemes}
\begin{figure} [h]
 \centerline{\includegraphics[width=3.5in]{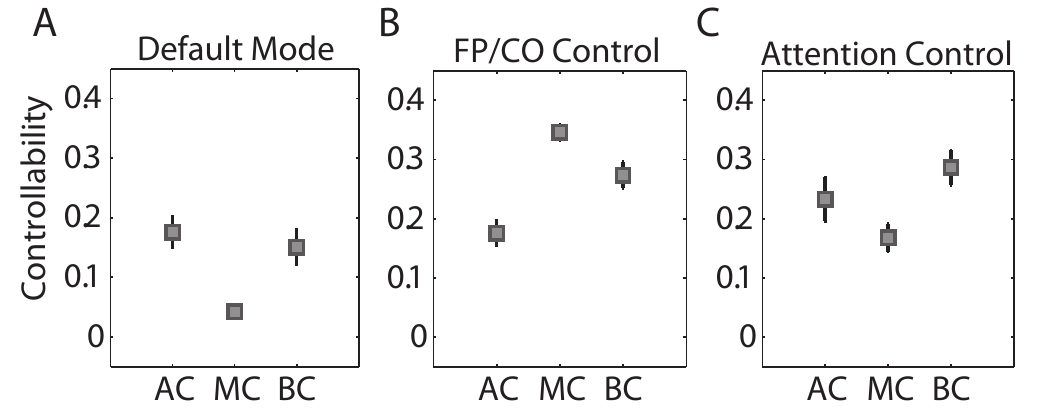}}
\caption[Differential Recruitment of Cognitive Systems to Network Control]{\textbf{Differential Recruitment of Cognitive Systems to Network Control} Average controllability (AC), modal controllability (MC), and boundary controllability (BC) hubs are differentially located in default mode \emph{(A)}, fronto-parietal and cingulo-opercular cognitive control \emph{(B)}, and attentional control \emph{(C)} systems. Values are averaged over the 3 replicates for each individual; error bars indicate standard deviation of the mean over subjects. \label{dsys_supmat}}
 \end{figure}

There is currently no accepted weighting scheme for constructing anatomical networks from diffusion imaging data. Weighting connections between ROIs based on the number of streamlines connecting them (as estimated by diffusion tractography algorithms) is the most commonly utilized scheme. However, it has been argued that these estimates can be biased by variation in the sizes of the regions under study \cite{Hagmann2008}: large regions may have a higher probability of displays more streamlines than smaller regions. While this potential bias does not appear to drastically alter large-scale topological properties of anatomical networks, its local effects are not well characterized \cite{Bassett2011}. Our results, based on the number of streamlines, are unlikely to be affected by this potential bias for one key reason: the Lausanne atlas family purposefully attempts to equalize region size, particularly in the higher scales \cite{Cammoun2012}. Nevertheless, to confirm that our results were robust to an alternative weighting scheme that accounts for region size, we divided each $ij^{th}$ element in the adjacency matrix $\mathbf{A}$ in Scale 125 ($N$=234) by the sum of the sizes of the two regions that it connects to create an alternative adjacency matrix $\mathbf{A}^{\prime}$. We then computed the controllability diagnostics and rank degree of $\mathbf{A}^{\prime}$ for each scan. Similar to our results obtains with the original weighting scheme, we observed that (i) the mean average controllability across the scans is strongly and positively correlated with mean rank degree ($r=0.97$, $p=1.43 \times 10^{-150}$), (ii) the mean modal controllability across the scans is strongly and negatively correlated with mean rank degree ($r=-0.96$, $p=2.51 \times 10^{-130}$), and (iii) the mean boundary controllability is not significantly correlated with mean rank degree ($r=-0.01$, $p=0.32$). Furthermore, the three network controllability diagnostics are again differentially recruited to known cognitive systems in the same manner as they were for the original weighting scheme (Compare Figure \ref{dsys_supmat} to Figure 4 in the main manuscript). Together, these findings indicate that the results reported in the main manuscript are robust to variations in weighting scheme that include a correction for region size.

\section*{Supplementary Discussion}

\subsection{Interpretations Dependent on Model Assumptions}
\addcontentsline{toc}{subsection}{Interpretations Dependent on Model Assumptions}
Network controllability differs significantly from the static graph theoretical approaches that are increasingly used in studies of human brain connectivity. Network controllability models brain dynamics based on two features: (i) a structural connectivity matrix and (ii) an equation of state defining the dynamics that occur on top of that structure. The theoretical predictions of network controllability diagnostics are therefore dependent on the accuracy of these two features. Here we utilize state-of-the art DSI imaging techniques \cite{Wedeen2005} and tractography reconstruction algorithms \cite{Yeh2011} to estimate white matter streamlines from the medial to lateral surfaces, and to distinguish their crossings \cite{Cieslak2014}. An underlying assumption assumption of this approach is that the number of streamlines is proportional to the strength of structural connectivity; this assumption has important exceptions but is most reasonable for cortico-cortical control, which is the primary area of investigation here (see SI for results from alternative weighting schemes). The equation of state that we utilize is based on extensive prior work demonstrating its utility in predicting resting state functional connectivity \cite{Honey2009} and in providing similar brain dynamics to more complicated models \cite{Galan2008}. Nevertheless, this model is simple, and our interpretations are dependent on its assumptions.

\section*{Supplemental Methods}
\subsection{Correlation Between Degree and Average Controllability}
\addcontentsline{toc}{subsection}{Correlation Between Degree and Average Controllability} In the main manuscript, we describe a strong correlation between node degree and
average controllability for the networks that we study. Here we provide a possible explanation for this effect. Due to the property $\mathrm{Trace}(\mathbf{ABC}) =
\mathrm{Trace}(\mathbf{BCA})$, the average controllability with a single
control node $j$ equals the $j$-th diagonal elements of
$(\mathbf{I}-\mathbf{A}^2)^{-1}$. Since $A$ is stable, a first order
approximation yields
\begin{equation}
(\mathbf{I}-\mathbf{A}^2)^{-1}\approx \mathbf{I} + \mathbf{A}^2 ,
\end{equation}
and the $j$-th diagonal element of $(\mathbf{I}-\mathbf{A}^2)^{-1}$ is
$1+\sum_{i=1}^{N}A_{ij}^2$. Since the degree of the $j$-th node equals
$d_j = \sum_{i=1}^{N} A_{ij}$, a positive correlation between node
degree and average controllability is mathematically expected in the networks that we study here.

\subsection{Lower Bound on the Largest Eigenvalue of the
  Controllability Gramian}
\addcontentsline{toc}{subsection}{Lower Bound on the Largest Eigenvalue of the
  Controllability Gramian}
In the main manuscript, we show that the smallest eigenvalue of the
controllability Gramian is in fact much smaller than its largest
counterpart. In fact, the largest eigenvalue of the controllability
Gramian is lower bounded by $1$. To see this, notice that
\begin{equation}
\lambda_{\max}(W_{\mathcal{K}}) = \lambda_{\max} \left( \sum_{\tau =
    0}^\infty \mathbf{A}^\tau
  \mathbf{B_{\mathcal{K}}}\mathbf{B_{\mathcal{K}}^\transpose} \mathbf{A}^\tau
\right) \ge \lambda_{\max} \left( \sum_{\tau =
    0}^0 \mathbf{A}^\tau
  \mathbf{B_{\mathcal{K}}}\mathbf{B_{\mathcal{K}}^\transpose} \mathbf{A}^\tau
\right) =  \lambda_{\max}(\mathbf{B_{\mathcal{K}}}\mathbf{B_{\mathcal{K}}^\transpose}) = 1,
\end{equation}
where the inequality follows from the fact that $\mathbf{A}^\tau
\mathbf{B_{\mathcal{K}}}\mathbf{B_{\mathcal{K}}^\transpose}
\mathbf{A}^\tau$ is positive semi-definite for all $\tau$.

\subsection{Additional Algorithmic Details for Boundary Control Method}
\addcontentsline{toc}{subsection}{Additional Algorithmic Details for Boundary Control Method}

In the main manuscript, we briefly describe our method for detecting boundary control points. This method is largely based on the algorithm proposed in \cite{FP-SZ-FB:13q}. However, for the application to brain networks derived from diffusion tractography, we have made two important modifications to more accurately estimate the initial partition and constrain the boundary point criteria as described in detail below.

\textbf{Initial Partition} The first modification concerns the definitions of the first level subnetworks for which we compute a two-partition based on the Fiedler eigenvector. In initial work, Pasqualetti et al. \cite{FP-SZ-FB:13q} suggest computing the Fiedler eigenvector of the adjacency matrix to create first level subnetworks defined by a two-partition. In contrast, we define this first level of subnetworks as composed of network communities, identified by maximizing the modularity quality function \cite{Newman2006} using a Louvain-like \cite{Blondel2008} locally greedy algorithm \cite{Jutla2011}. Our choice is based on extensive recent literature demonstrating that the brain is composed of many subnetworks (not just 2) \cite{Meunier2010,Bassett2013}, which can be extracted using modularity maximization approaches \cite{Meunier2009,Chen2008,Bassett2011b}, and which correspond to sets of brain areas performing related functions \cite{Power2012,Chen2008}.

The modularity quality function provides an estimate of the quality of a hard partition of the $N \times N$ adjacency matrix $\mathbf{A}$ into network communities (whereby each brain region is assigned to exactly one network community) \cite{NG2004,markfast,Newman2006,Porter2009,Fortunato2010}
\begin{equation}\label{one}
	Q_{0} = \sum_{ij} [A_{ij} - \gamma P_{ij}] \delta(g_{i},g_{j})\,,
\end{equation}
where brain region $i$ is assigned to community $g_{i}$, brain region $j$ is assigned to community $g_{j}$, $\delta(g_{i},g_{j})=1$ if $g_{i} = g_{j}$ and it equals $0$ otherwise, $\gamma$ is a structural resolution parameter, and $P_{ij}$ is the expected weight of the edge connecting node $i$ and node $j$ under a specified null model. Maximization of $Q_{0}$ yields a hard partition of a network into communities such that the total edge weight inside of communities is as large as possible (relative to the null model and subject to the limitations of the employed computational heuristics, as optimizing $Q_{0}$ is NP-hard \cite{Porter2009,Fortunato2010,Brandes2008}).

Because the modularity quality function has many near-degeneracies, it is important to perform the optimization algorithm multiple times \cite{Good2010}. We perform 100 optimizations of the Louvain-like locally greedy algorithm \cite{Jutla2011} for each adjacency matrix corresponding to a single scan. To distill a single representative partition, we create a consensus partition from these 100 optimizations based on statistical comparison to an appropriate null model \cite{Bassett2013}.

In a final consideration, we choose a value for the structural resolution parameter $\gamma$. The choice $\gamma  = 1$ is very common, but it is important to consider multiple values of $\gamma$ to examine community structure at multiple scales \cite{rb2006,Porter2009,Onnela2011}. Indeed, recent work has demonstrated that in some networks, a structural resolution parameter value that accurately captures the underlying community structure can be identified by the $\gamma$ value at which the 100 optimizations produce similar partitions \cite{Bassett2013}. To quantitatively estimate similarity in partitions, we adopt the $z$-score of the Rand coefficient \cite{Traud2010}. For each pair of partitions $\alpha$ and $\beta$, we calculate the Rand $z$-score in terms of the total number of pairs of nodes in the network $M$, the number of pairs $M_{\alpha}$ that are in the same community in partition $\alpha$, the number of pairs $M_{\beta}$ that are in the same community in partition $\beta$, and the number of pairs of nodes $w_{\alpha \beta}$ that are assigned to the same community both in partition $\alpha$ and in partition $\beta$. The $z$-score of the Rand coefficient comparing these two partitions is
\begin{equation}
	z_{\alpha\beta} = \frac{1}{\sigma_{w_{\alpha \beta}}} w_{\alpha \beta}-\frac{M_{\alpha}M_{\beta}}{M}\,,
\end{equation}
where $\sigma_{w_{\alpha \beta}}$ is the standard deviation of $w_{\alpha \beta}$. Let the \emph{mean partition similarity} denote the mean value of $z_{\alpha \beta}$ over all possible partition pairs for $\alpha \neq
\beta$. Let the \emph{variance of the partition similarity} denote the variance of $z_{\alpha \beta}$ over all possible partition pairs for $\alpha \neq \beta$.

Empirically, we calculated a group adjacency matrix by averaging the adjacency matrices of all subjects and scans. We optimized the modularity quality function 100 times and we computed the mean and variance of the partition similarity for a range of $\gamma$ values and for all 5 spatial resolutions. Across all atlases, we observed that the mean partition similarity was high and the variance of the partition similarity was low for values of $\gamma$ ranging between $1.5$ and $2$. For Scale 125 (the atlas for which we report results in the main manuscript), we observed a maximum mean partition similarity and a minimum variance of partition similarity at $\gamma=1.6$. We therefore chose to set $\gamma = 1.6$ for the remainder of the analysis in this study.

\begin{figure}
 \centerline{\includegraphics[width=4in]{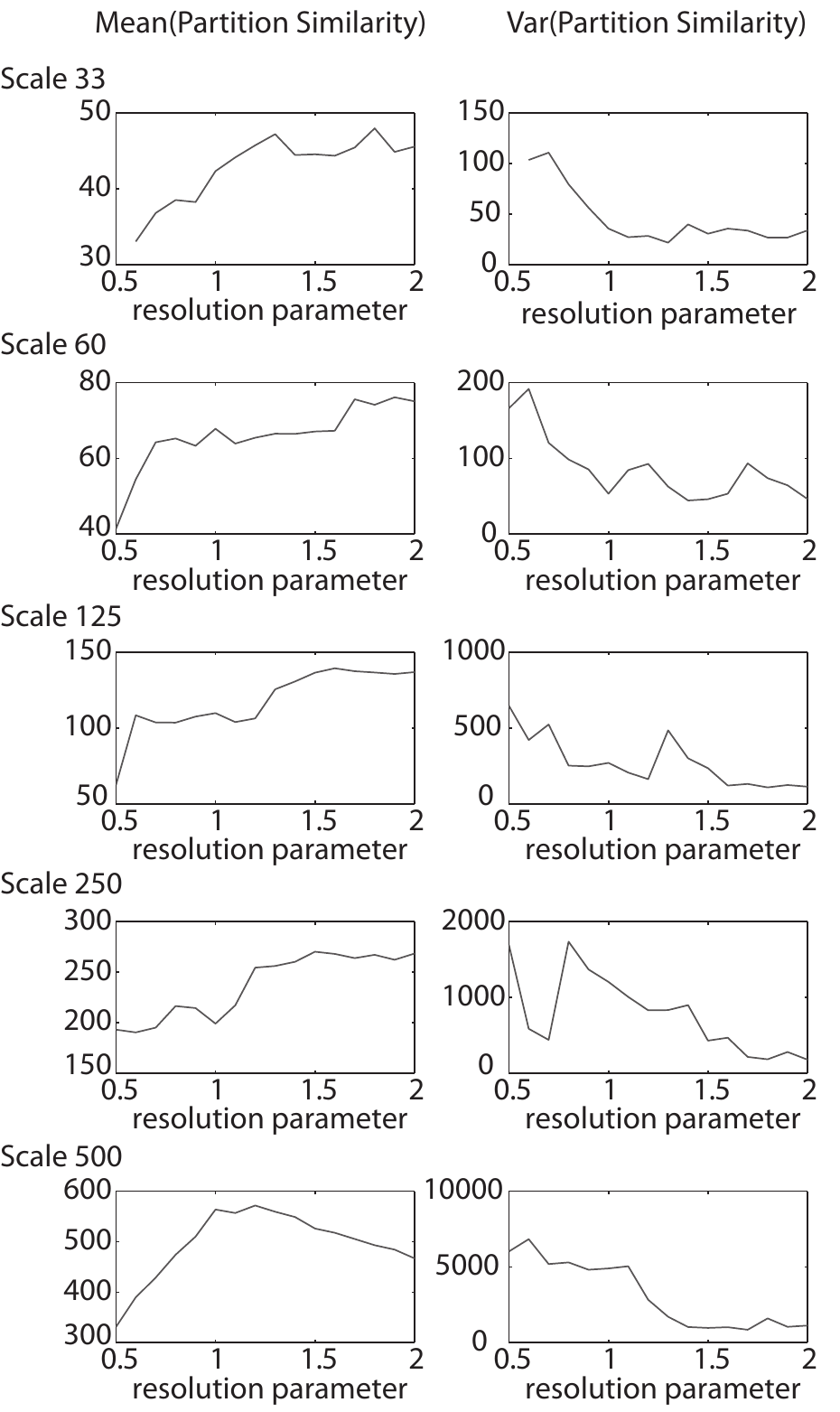}}
\caption[Partition Similarity As a Function of the Resolution Parameter]{\textbf{Partition Similarity As a Function of the Resolution Parameter} Mean \emph{(left)} and variance \emph{(right)} of the partition similarity estimated using the $z$-score of the Rand coefficient as a function of the structural resolution parameter $\gamma$, varies from $0.5$ to $2$ in increments of $0.1$, for the 5 spatial scales of the Lausanne atlas \cite{Hagmann2008} \emph{(rows)}. \label{gamma}}
 \end{figure}

\textbf{Boundary Point Criteria} The second modification concerns the definition of a boundary point. After calculating the Fiedler eigenvector of a subnetwork to determine a partition of the subnetwork into two communities, we must identify ``boundary points'', which are nodes that contain connections to both communities. In the original work by Pasqualetti and colleagues, it was suggested that a boundary point was a node with any number of connections to both communities. However, in weighted brain networks we suggest that a more stringent definition is more appropriate for the following reason: practically all nodes in the brain have non-zero weighted connections to both identified communities. Therefore, we instead set a threshold ratio $\rho$ to identify boundary points. Considering the adaptivity to the local measure, we set a threshold ratio $\rho$ instead of a global threshold value. In detail, for a network $G =(V, E)$ with partition $P = (V_1, \cdots, V_n)$, a node $i\in V_k$ is called a boundary node if
\begin{equation}
\sum_{l\neq k}a_{kl} \geq \rho\cdot\max(A)
\end{equation}
where $A$ is the adjacency matrix. Here, $\max(A)$ can be replaced with other statistics and $\rho$ needs to be chosen carefully. If $\rho$ is too small, there will be no effect and the algorithm tends to add the total subnetwork as the set of boundary points. If $\rho$ is too large, there will be only a few points recognized as the boundary points.

\begin{figure}
 \centerline{\includegraphics[width=3.5in]{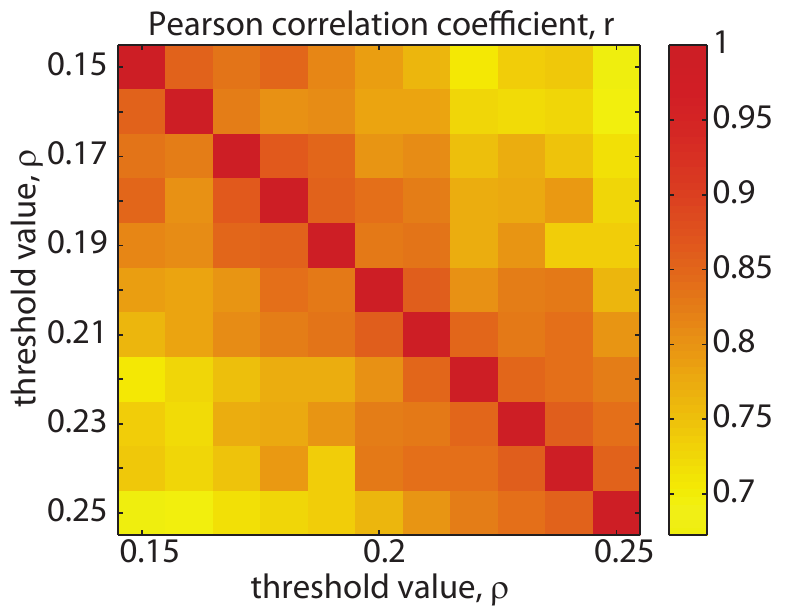}}
\caption[Effect of Boundary Point Criteria Threshold]{\textbf{Effect of Boundary Point Criteria Threshold}  Color indicates Pearson correlation coefficient, $r$, between the vectors of boundary controllability values estimated for pairs of $\rho$ values in the range $0.15-0.25$ in increments of $0.01$. \label{rho}}
 \end{figure}

In the results described in the main manuscript, we set the threshold ratio to $\rho = 0.2$. To determine whether our results are robust to this choice, we calculate boundary controllability values across all regions in the Scale 125 atlas, for each scan using $\rho$ values that vary between $0.15$ and $0.25$ in increments of $0.01$. We then asked how similar regional control values were for different choices of $\rho$. Specifically, for any pair of $\rho$ values, we computed the Pearson correlation coefficient between the vectors of regional control values for the two $\rho$ values. We show the results of this analysis in Fig.~\ref{rho}. We observe that the boundary control values are highly similar across choices of $\rho$ (minimum Pearson correlation approximately $0.68$, corresponding to a $p=0$, indicating that our results are robust to small variation in the boundary point criteria threshold.

\textbf{Final Algorithm}
Thus, the final algorithm used in the calculation of boundary controllability in this paper can be summarized as follows. We begin with the application of a community detection method to the brain network to extract a partition of brain regions into network communities. We then recursively apply a Fiedler bipartition to add boundary nodes within communities, with the goal of improving the local controllability of the network. At each stage of the algorithm, we define the boundary nodes of the network as the nodes that maintain edges to nodes in other communities. Algorithmically, we can write:

\begin{algorithm}[H]
 \SetAlgoLined
 \KwData{Network $G = (V, E)$ with adjacency matrix $A = (a_{ij})$, Number of control nodes $m$, threshold ratio $\rho$;}
 \KwResult{Control Nodes Index Set $\mathcal{K}$; }
 Define an empty set of control nodes $\mathcal{K} = \emptyset$\;
 Initialize the partition $\mathcal{P}$ with the result of a community detection algorithm and initialize the boundary nodes set $\mathbf{B} = \emptyset$\;
 Add the boundary points of the initial partition\;
 \While{$|\mathcal{K}| < m$}{
  Select least controllable community $l = \arg\min\{\lambda_{min}(W_{i,\infty}), i = 1, ..., |\mathcal{P}|\}$\;
  Compute Fiedler two-partition $P_f$ of $l$-th community\;
  Compute boundary nodes $B_f$ of $P_f$ with the given threshold ratio $\rho$\;
  Update partition $\mathcal{P}$ with $P_f$\;
  Update control nodes with boundary nodes $\mathcal{K} = \mathcal{K}\cup B_f$\;
  }
   \Return $\mathcal{K}$.
 \caption{Algorithm for the Selection of Boundary Control Nodes}\label{alg1}
\end{algorithm}

\subsection{Association of Brain Regions to Cognitive Systems}
\addcontentsline{toc}{subsection}{Association of Brain Regions to Cognitive Systems}

To examine the relationship between controllability diagnostics and cognitive systems, we developed a map of brain areas to a set of cognitive systems previously defined in the literature: the fronto-parietal, cingulo-opercular, dorsal attention, ventral attention, default mode, motor and somatosensory, auditory, visual, subcortical systems \cite{Power2012}. Such a mapping was inspired by a recent paper from Power et al. (2012) who associated 264 brain areas to these cognitive systems, defined by a clustering technique applied to functional brain networks \cite{Power2012}. Similar to previous work \cite{Power2012}, our association of areas to systems is a gross approximation and it should not be interpreted as indicating that areas have single functions. We use this association only as a pragmatic means to assess whether controllability diagnostics are differentially identified in distributed neural circuits.

The 234 areas examined in the main manuscript were drawn from 42 cortical structures. Here we associate these 42 structures to the set of 9 cognitive systems:
\begin{itemize}
    \item \textbf{Lateral Orbitofrontal} In the Power et al. (2012) decomposition, portions of lateral orbitofrontal cortex (or BA 47) are assigned to default mode, salience, and ventral attention systems. To choose a single association for this region, we turned to the wider literature. In a recent meta-analysis, Zald and colleagues examined the role of medial and lateral orbitofrontal cortex in widespread functional networks \cite{Zald2014}. The lateral orbitofrontal cortex showed co-activations with prefrontal regions and areas involved in cognitive functions including language and memory but not with areas of the default mode, autonomic, and limbic systems. Rothkirch et al. (2012) similarly demonstrated that lateral orbitofrontal cortex appears to be modulated by implicit motivational value, rather than salience \cite{Rothkirch2012}, arguing against its inclusion in the salience system. Anderson and colleagues suggest that lateral orbitofrontal cortex provides a specificity in top-down control of attention in collaboration with dorsolateral prefrontal cortex \cite{Anderson2010}. Cognitive system assignment: ``Ventral Attention''.
    \item \textbf{Pars Orbitalis} In the Power et al. (2012) decomposition, portions of pars orbitalis (or BA 47) are assigned to default mode, salience, and ventral attention systems. To choose a single association for this region, we turned to the wider literature. The pars orbitalis is a part of the ventrolateral prefrontal cortex, and is known to play a role in cognitive control processes \cite{Xiang2010}, particularly in conflict adaptation \cite{Egner2011}, inhibition \cite{Enriquez2013}, which differ significantly from those enabled by the fronto-parietal network \cite{Elton2014}. Cognitive system assignment: ``Cingulo-Opercular''.
    \item \textbf{Frontal Pole} In this parcellation scheme, the frontal pole corresponds to portions of BA 9 and 10. These areas form hubs of the fronto-parietal cognitive control system \cite{Tu2013}. Cognitive system assignment: ``Fronto-parietal''.
    \item \textbf{Medial Orbitofrontal}. The medial frontal cortex is one of the key hubs of the fronto-parietal network \cite{Tu2013,Stern2012}. Cognitive system assignment: ``Fronto-parietal''.
    \item \textbf{Pars Triangularis} In this parcellation scheme, the pars triangularis corresponds to portions of BA 45, and therefore maps to the fronto-parietal cognitive control system \cite{Elton2014}. Cognitive system assignment: ``Fronto-parietal''.
    \item \textbf{Pars Opercularis} The pars of opercularis (corresponding roughly to BA 44) forms a hub of the cingulo-opercular cognitive control system \cite{Elton2014}. Cognitive system assignment: ``Cingulo-Opercular''.
    \item \textbf{Rostral Middle Frontal} The rostral middle frontal cortex, corresponding roughly to BA 10, forms a hub of the cingulo-opercular cognitive control system \cite{Elton2014}. Cognitive system assignment: ``Cingulo-Opercular''.
    \item \textbf{Superior Frontal}. In the Power et al. (2012) decomposition, portions of the superior frontal cortex are predominantly affiliated with the default mode system, consistent with previous literature \cite{Xu2013,Lukoshe2013,Sun2013,Fang2013}. Cognitive system assignment: ``Default Mode''.
    \item \textbf{Caudal Middle Frontal} The caudal middle frontal cortex is a prefrontal cortical structure broadly associated with executive function \cite{Marques2014,Lopez2011}, top-down control \cite{Durazzo2011}, and secondary motor processes \cite{Verstraete2011,Duffield2013}. Cognitive system assignment: ``Fronto-parietal''.
    \item \textbf{Precentral} The precentral cortex is part of the somatosensory system. Cognitive system assignment: ``Somatosensory''.
    \item \textbf{Paracentral} The paracentral cortex is part of the somatosensory system. Cognitive system assignment: ``Somatosensory''.
    \item \textbf{Rostral Anterior Cingulate} The anterior cingulate is a hub of the cingulo-opercular network \cite{Ninaus2013,Gratton2013,Vaden2013,Becerril2013,Sestieri2014,Tu2012}. Cognitive system assignment: ``Cingulo-Opercular''.
    \item \textbf{Caudal Anterior Cingulate} The anterior cingulate is a hub of the cingulo-opercular network \cite{Ninaus2013,Gratton2013,Vaden2013,Becerril2013,Sestieri2014,Tu2012}. Cognitive system assignment: ``Cingulo-Opercular''.
    \item \textbf{Posterior Cingulate}. The posterior cingulate is a known hub of the default mode system \cite{Stern2012,Khalsa2013,Leech2014}. Cognitive system assignment: ``Default Mode''.
    \item \textbf{Isthmus Cingulate} The isthmus cingulate is thought to be a hub of the default mode system \cite{Zhu2013} and of the limbic system \cite{Leow2013}. Cognitive system assignment: ``Default Mode''.
    \item \textbf{Post Central} The postcentral cortex is part of the somatosensory system. Cognitive system assignment: ``Somatosensory''.
    \item \textbf{Supramarginal} The supramarginal gyrus appears to play a role in the dorsal \cite{Schmidt2013} and ventral \cite{Burianova2012} attention networks, and executive function more broadly \cite{Yin2012,Vasconcelos2014}. In the Power et al. (2012) decomposition, this area was assigned to the cingulo-opercular system \cite{Power2012}. Cognitive system assignment: ``Cingulo-Opercular''.
    \item \textbf{Superior Parietal} The superior parietal cortex plays a role in both the dorsal attention system \cite{Xu2014,Sestieri2013} and the somatosensory-motor system \cite{Fabbri2014}. Cognitive system assignment: ``Dorsal Attention''.
    \item \textbf{Inferior Parietal}. The inferior parietal cortex is one of the key hubs of the fronto-parietal network \cite{Tu2013,Stern2012}. Cognitive system assignment: ``Fronto-parietal''.
    \item \textbf{Precuneus} The precuneus is a hub of the default mode system \cite{Cavanna2006,Sheline2013}. Cognitive system assignment: ``Default Mode''.
    \item \textbf{Cuneus} The cuneus is a part of the visual system \cite{Xu2014,Delli2014,Collignon2013}. Cognitive system assignment: ``Visual''.
    \item \textbf{Pericalcarine} The pericalcarine is a part of the visual system \cite{Gaetz2012,Delli2014}. Cognitive system assignment: ``Visual''.
    \item \textbf{Lateral Occipital} The lateral occipital cortex is a part of the visual system \cite{Bedny2012}. Cognitive system assignment: ``Visual''.
    \item \textbf{Lingual} The lingual gyrus is a part of the visual system \cite{Delli2014,Boldt2014}. Cognitive system assignment: ``Visual''.
    \item \textbf{Fusiform} The lingual gyrus is a part of the visual system \cite{Xu2014}. Cognitive system assignment: ``Visual''.
    \item \textbf{Parahippocampal} The parahippocampal cortex has been associated with many cognitive processes including visuospatial processing and episodic memory \cite{Aminoff2013}. Cognitive system assignment: ``Other''.
    \item \textbf{Entorhinal cortex} The entorhinal cortex encodes visual information \cite{Killian2012}. Cognitive system assignment: ``Visual''.
    \item \textbf{Temporal Pole} The temporal pole plays a role in language processing, including naming \cite{Semenza2011}, and in social and emotional processing \cite{Olson2007}. Cognitive system assignment: ``Other''.
    \item \textbf{Inferior Temporal} The inferior temporal cortex is associated with visual processing \cite{Hirabayashi2014}, emotion perception of visual objects \cite{Sabatinelli2011}, and shape recognition \cite{Tompa2010}. Cognitive system assignment: ``Visual''.
    \item \textbf{Middle Temporal} The middle temporal cortex is associated with cognitive control processes \cite{Noonan2013}, theory of mind \cite{Rodrigo2014}, and social cognition \cite{Hayashi2014}. Cognitive system assignment: ``Other''.
    \item \textbf{Bank of the Superior Temporal Sulcus} The bank of the superior temporal sulcus forms a part of the early cortical auditory network \cite{Kilian2011}. Cognitive system assignment: ``Auditory''.
    \item \textbf{Superior Temporal} The superior temporal cortex forms a part of the auditory system \cite{Woods2009}. Cognitive system assignment: ``Auditory''.
    \item \textbf{Transverse Temporal} The transverse temporal cortex forms a part of the auditory system \cite{Simon2013}. Cognitive system assignment: ``Auditory''.
    \item \textbf{Insula}. The insula is one of the key hubs of the fronto-parietal network \cite{Tu2013,Stern2012}. Cognitive system assignment: ``Fronto-parietal''.
    \item \textbf{Thalamus}. Cognitive system assignment: ``Subcortical''.
    \item \textbf{Caudate}. Cognitive system assignment: ``Subcortical''.
    \item \textbf{Putamen}. Cognitive system assignment: ``Subcortical''.
    \item \textbf{Pallidum}. Cognitive system assignment: ``Subcortical''.
    \item \textbf{Nucleus Accumbens}. Cognitive system assignment: ``Subcortical''.
    \item \textbf{Hippocampus}. Cognitive system assignment: ``Subcortical''.
    \item \textbf{Amygdala}. Cognitive system assignment: ``Subcortical''.
    \item \textbf{Brainstem}. Cognitive system assignment: ``Other''.
\end{itemize}

\newpage
\bibliographystyle{naturemag}
\bibliography{./bibfile}









 \end{document}